\documentclass[letter,12pt]{article}
\usepackage[paper=letterpaper,margin=1in]{geometry}

\usepackage{amsmath,amssymb,amsfonts,epsfig,cite,setspace,bigstrut,longtable,array}
\usepackage{hyperref}
\usepackage{amsthm}

\pdfoutput=1

\begin{document}

%%%%%%%%%%%%%%%%%%%%%%%%%%%%%%%%%%%%%%%%%%%%%%%%%%%%%%%%%%%%%%%%%%

\vskip 0.25in

\newcommand{\todo}[1]{{\bf ?????!!!! #1 ?????!!!!}\marginpar{$\Longleftarrow$}}
\newcommand{\fref}[1]{Figure~\ref{#1}}
\newcommand{\tref}[1]{Table~\ref{#1}}
\newcommand{\sref}[1]{\S~\ref{#1}}
\newcommand{\nn}{\nonumber}
\newcommand{\tr}{\mathop{\rm Tr}}
\newcommand{\comment}[1]{}

\newcommand{\cM}{{\cal M}}
\newcommand{\cA}{{\cal A}}
\newcommand{\cW}{{\cal W}}
\newcommand{\cN}{{\cal N}}
\newcommand{\cH}{{\cal H}}
\newcommand{\cK}{{\cal K}}
\newcommand{\cZ}{{\cal Z}}
\newcommand{\cO}{{\cal O}}
\newcommand{\cB}{{\cal B}}
\newcommand{\cC}{{\cal C}}
\newcommand{\cD}{{\cal D}}
\newcommand{\cE}{{\cal E}}
\newcommand{\cF}{{\cal F}}
\newcommand{\cX}{{\cal X}}
\newcommand{\IA}{\mathbb{A}}
\newcommand{\IP}{\mathbb{P}}
\newcommand{\IQ}{\mathbb{Q}}
\newcommand{\IH}{\mathbb{H}}
\newcommand{\IR}{\mathbb{R}}
\newcommand{\IC}{\mathbb{C}}
\newcommand{\IF}{\mathbb{F}}
\newcommand{\IV}{\mathbb{V}}
\newcommand{\II}{\mathbb{I}}
\newcommand{\IE}{\mathbb{E}}
\newcommand{\IZ}{\mathbb{Z}}
\newcommand{\re}{{\rm Re}}
\newcommand{\im}{{\rm Im}}
\newcommand{\ch}{{\rm ch}}
\newcommand{\sym}{{\rm Sym}}

\newcommand{\tmat}[1]{{\tiny \left(\begin{matrix} #1 \end{matrix}\right)}}
\newcommand{\mat}[1]{\left(\begin{matrix} #1 \end{matrix}\right)}
\newcommand{\diff}[2]{\frac{\partial #1}{\partial #2}}
\newcommand{\gen}[1]{\langle #1 \rangle}
\newcommand{\ket}[1]{| #1 \rangle}
\newcommand{\jacobi}[2]{\left(\frac{#1}{#2}\right)}

\newcommand{\drawsquare}[2]{\hbox{%
\rule{#2pt}{#1pt}\hskip-#2pt%  left vertical
\rule{#1pt}{#2pt}\hskip-#1pt%  lower horizontal
\rule[#1pt]{#1pt}{#2pt}}\rule[#1pt]{#2pt}{#2pt}\hskip-#2pt%  upper horizontal
\rule{#2pt}{#1pt}}% right vertical
\newcommand{\fund}{\raisebox{-.5pt}{\drawsquare{6.5}{0.4}}}
\newcommand{\antifund}{\overline{\fund}}

\newtheorem*{theorem}{\bf THEOREM}
\def\thetheorem{\thesection.\arabic{theorem}}
\newtheorem*{proposition}{\bf PROPOSITION}
\def\thetheorem{\thesection.\arabic{proposition}}
\newtheorem*{conjecture}{\bf CONJECTURE}
\def\thetheorem{\thesection.\arabic{conjecture}}
\newtheorem*{definition}{\bf DEFINITION}
\def\thetheorem{\thesection.\arabic{definition}}

\def\theequation{\thesection.\arabic{equation}}
\newcommand{\setall}{\setcounter{equation}{0}
        \setcounter{theorem}{0}}
\newcommand{\setequation}{\setcounter{equation}{0}}
\newcommand\blfootnote[1]{%
  \begingroup
  \renewcommand\thefootnote{}\footnote{#1}%
  \addtocounter{footnote}{-1}%
  \endgroup
}

~\\
\vskip 1cm

\begin{center}
{\Large \bf Extremal Bundles on Calabi-Yau Threefolds}
\medskip

\vspace{.4cm}
\centerline{
{\large Peng Gao}$^1$,
{\large Yang-Hui He}$^2$, 
{\large Shing-Tung Yau}$^1$
}
\vspace*{3.0ex}
{\it
\vspace*{1.5ex}
{\small
{
${}^{1}$
Department of Mathematics, Harvard University, Cambridge MA 02138, USA;\\
Taida Institute for Mathematical Science, National Taiwan University, Taipei, Taiwan\\
}
{${}^{2}$ 
Department of Mathematics, City University, London, EC1V 0HB, UK; \\
School of Physics, NanKai University, Tianjin, 300071, P.R.~China; \\
Merton College, University of Oxford, OX14JD, UK\\
}
\blfootnote{
penggao@math.harvard.edu, \quad
hey@maths.ox.ac.uk, \quad
yau@math.harvard.edu
}
}
}
\end{center}

\vspace*{4.0ex}
\centerline{\textbf{Abstract}} \bigskip
We study constructions of stable holomorphic vector bundles on Calabi-Yau threefolds, especially those with exact anomaly cancellation which we call extremal. By going through the known databases we find that such examples are rare in general and can be ruled out for the spectral cover construction for all elliptic threefolds. We then introduce a generalized version of Hartshorne-Serre construction and use it to yield extremal bundles of general ranks and study their geometry. In light of this probing the geometry of the space of stable vector bundles, we revisit the DRY conjecture on stable reflexive sheaves while focusing on the distribution of Chern numbers to use both theoretical and statistical ideas to provide evidence for DRY.

\newpage

\tableofcontents

%%%%%%%%%%%%%%%%%%
%================
\section{Introduction}

The theorems of Donaldson, Uhlenbeck and Yau (DUY) \cite{don,UY} connects the study of the Hermitian-Yang-Mills connections on complex algebraic varieties, a difficult set of non-linear partial differential equations, to that of (poly-)stable sheaves, furnishing us with an algebraic handle. 
This correspondence has greatly facilitated the study of these objects in physical theories, especially in context of the so called Heterotic string where the stable holomorphic vector bundles directly determines the low energy supersymmetric quantum field theory particle spectrum and their couplings. 
Understanding the structure of Chern and secondary characteristic classes is of great importance for the model building efforts in Heterotic string theory \cite{Anderson:2011ty}.  It is within this context that we will explore the holomorphic vector bundles on Calabi-Yau threefolds, their topological invariants and numerical relations.  

Purely theoretically the threefold case is also of great interest, particularly due to fundamental difficulties involved in understanding the structure of (compactification) of the moduli space of (semi-)stable holomorphic vector bundles in this dimension. In complex two dimensions, the analytical convergence required for compactification of the moduli spaces is guaranteed by the so-called Uhlenbeck compactness \cite{Uhlenbeck}.  In terms of algebraic geometry, the corresponding construction arises from Mumford-Takemoto and Gieseker's stability construction and a purely algebraic embedding of the moduli space homeomorphic to Uhlenbeck compacitification was given in \cite{LiJun}. The parallel construction in complex dimension three is so far an open problem. In physics terms this may be related to the fact that gauge theories in (real) dimensions above four are generally nonrenormalizable. 

One very natural approach to resolve these issues is suggested by the physical role that the holomorphic vector bundles play in the heterotic string theory, namely they furnish the space of supersymmetric vacua for the full string theory. In the related context of supersymmetric vacua of type II string theories (see e.g. \cite{Denef:2004ze}), a particular way to understand these vacua is as non-degenerate critical points of a naturally associated functional \cite{Ashok:2003gk}, the superpotential, parametrized by physical coupling constants which are the counterparts of the information encoded in the mathematical deformation theory. For further developments of this approach in the mathematics literature, see for example \cite{Douglas:2004zu}.

For threefolds, there is by now a rather well motivated mathematical theory about the enumeration of stable sheaves on complex threefolds \cite{DT}. In cases of rank higher than one, the case of more physical relevance for the heterotic theory, the theory is only well-defined for Calabi-Yau threefold case (i.e. trivial canonical bundle). For mathematical backgrounds related to this and also the question regarding virtual fundamental cycles see for example \cite{LiWu}.    
The superpotential whose critical points give rise to holomorphic vector bundles \footnote{Strictly speaking the stability property is not encoded in the superpotential, whose variation gives rise to holomorphic Yang-Mills equations, rather than slope-stable Hermitian Yang-Mills equations.}, is the so-called holomorphic Chern-Simons action originally defined in \cite{Witten:1992fb} (see also \cite{Kachru:2000ih,Aganagic:2000gs,Lazaroiu:2001bz}).
An essential aspect of the difficulty for studying moduli space of vector bundles on threefolds comes from the fact that Hilbert schemes of (subschemes of) projective threefolds are non-reduced, hence singular locally everywhere\cite{BBS}. 

There have been recently important advances in the understanding of these moduli spaces. However, despite these powerful advances, we are still confronted with technical hurdles which render their role in the corresponding physical theories largely mysterious, especially in terms of compact Calabi-Yau varieties. Moreover, explicit expressions for the DT invariants, especially for bundles of higher rank on compact threefolds, are hard to obtain (see however \cite{Thomas,LQ}).
At a practical level, for model building one may be contented with the understanding of slope stability at a case by case level. Here, to prove stability one is typically entangled with the difficult combinatorial problem of finding the lattice of all sub-sheaves before checking their slope and verifying or ruling out stability. For higher rank vector bundles, this becomes a highly complex computational problem. 

With some hindsight, such complexity is perhaps not unexpected, given the similar situation in type II constructions \cite{DeWolfe:2004ns,Denef:2006ad}. In fact, it becomes somewhat pleasant in this light to consider the relation with the mathematical complexity of moduli spaces (Hilbert schemes) which is `unbounded' in terms of types of singularities that can occur there already in the case of algebraic curves \cite{Vakil} \footnote{This has been dubbed `Murphy's law' in that context. While in physics it is known as `Murray Gell-Mann's totalitarian principle' in quantum mechanics.}.

Generalizing known inequalities for holomorphic vector bundles on complex surfaces \cite{bogomolov}, in \cite{Douglas:2006jp}  the third author and collaborators employed the attractor flow arguments in string theory to arrive at a set of conjectured (sufficient) conditions -- the so-called DRY conditions -- for the existence of stable bundles on Calabi-Yau threefolds. Important new findings about the split attractor flow equations have meanwhile appeared \cite{Denef:2007vg}, and it seems interesting to relate the attractor flow equation to the (conjectured) holomorphic version of Floer theory \cite{Thomas}.
 
 The DRY conditions \cite{Douglas:2006jp}  are numerical relations among Chern characters of the hypothetical holomorphic vector bundle and the polarizing ample divisor(s). This makes it interesting to understand such numerical relations further based on known examples\footnote{Indeed, for a given Chern vector specifying the various Chern classes of a sheaf, the Donaldson-Thomas (DT) invariants \cite{DT},\cite{Thomas} serve a generating function for counting such (semi-)stable objects.}. 
A novel approach to this problem is the use of algorithmic methods for overcoming the combinatorial complexity. And the scope of the present paper is to make use of the algorithmic approach in addition to mathematical constructions, to shed some new light on the DRY conjecture. 
Readers interested mainly in the results can find a summary in the concluding Section \ref{concl}.

%%%%%%%%%%%%%%%%%%%%%%%%%%%
\section{Extremal Bundles and Chern Class Constraints}

The last decade has witnessed tremendous advances in computer algebra and algorithmic algebraic geometry, especially in the context of gauge theories \cite{comp-book} and applications to Calabi-Yau threefolds (cf.~\cite{He:2013epn}).
Building upon the now classic datasets of smooth Calabi-Yau threefolds, notably the earliest set of complete intersections \cite{Candelas:1987kf,Gagnon:1994ek} in products of projective spaces (CICY) and the impressive $\sim 10^9$ hypersurfaces in toric varieties \cite{Avram:1997rs} by Kreuzer-Skarke (KS), various databases of stable vector bundles have been compiled \cite{Anderson:2012yf,Anderson:2013xka,Donagi:2004ia,Gabella:2008id,Anderson:2007nc,Anderson:2008uw,He:2011rs,Blumenhagen:2011xn,Gao:2013pra}.
Thus armed, let us investigate some interesting features in the space of stable unitary bundles on a smooth Calabi-Yau threefold $X$, focusing especially on the statistics.

Of the classes of possible candidates, we will place our attention on $SU(n)$ bundles $V$, this not only for concreteness and specificity, but also is well motivated by physical reasons which we will shortly address.
Bundles of such type automatically satisfies the vanishing of the first Chern class because of the {\em special} unitary nature of the gauge group, viz., the vanishing of the trace of the field strength.
Next, we will place constraints on the second Chern class as well.
Indeed, there is a distinguished bundle to any smooth algebraic variety, namely its tangent bundle.
For a Calabi-Yau threefold $X$, that its tangent bundle $TX$ is of $SU(3)$ holonomy is one of the working definitions of the manifold.
Of particular interest to us is when our bundle $V$ has second Chern class {\it equal} to $\ch_2(TX)$, this is again not only for the sake of finiteness (unless we place a constraint on the second Chern class, there will typically be an infinite number of stable objects for some polarization), but also motivated by physics.

In particular, within the context of perturbative heterotic string theory, for bundles with second Chern class different from $\ch_2(TX)$, \footnote{Strictly speaking, agreement at the level of cohomology classes does not guarantee existence of a solution. }  one needs to `soak up' the anomaly by  introducing a nonvanishing $H$-flux, solving the Strominger system of \cite{Strominger:1986uh}. For mathematical progress in this direction, see the series of papers \cite{FuYau1,FuYau2,FuYau3,FuLiYau}. Often, in models which admit heterotic M5-branes, this difficulty is circumvented to some degree. 
We now introduce the notion of `extremality' and focus on bundles for which anomaly cancellation is exact. 

It is worth-while pointing out that in the smooth category more things are possible, and actual physical transitions may violate certain holomorphic quantities including such things as Chern numbers of subbundles with respect to the $E_8$ bundle, like those discussed here. But we will constrain ourselves to stay in the holomorphic/analytic category as is appropriate for discussing the configurations of the physical vacuum.

\begin{definition}
A stable vector bundle $V$ on a Calabi-Yau threefold $X$ is called {\bf extremal} {\rm if}
\begin{equation}\label{cherncons}
\begin{array}{ccl}
(1) & & \ch_1(V) = 0 ; \\
(2) & & \ch_2(TX) = \ch_2(V) . \\
%(3) & & \ch_3(V) = 3k \ , \mbox{ with } k \in \IZ \mbox{ dividing } \chi(X) \ .
\end{array}
\end{equation}
\end{definition}
The purpose of this paper is to study the space of (semi-)stable bundles, especially with an interest in finding those which are extremal. In this paper, we  take largely an ``experimental'' approach towards this question\footnote{
In later sections (especially in relation to the DRY conjecture), we also impose other physics motivated numerical constraints on the Chern classes mainly to to simplify our numerical work, but for most of the present paper we will focus on these two conditions, the vanishing of the first Chern class and the equality of the second to that of the tangent bundle, such bundles are called {\bf extremal} from now on.}. 

Indeed, on arguably the most famous Calabi-Yau threefold, the quintic hypersurface in $\IP^4$, such extremal bundles have been considered in \cite{Douglas:2004yv} for chirality change in the four dimensional heterotic string vacua.
We will shortly see the generalized constructions of this types to other large classes of Calabi-Yau threefolds as well as to other types of extremal bundle constructions.

%%%
\subsection{Physical Motivations}
As mentioned above, there are physical inspirations to consider extremal bundles; they come from string theory compactifications.
In the context of the heterotic string with $E_8$ gauge group \footnote{Let us for now not consider the hidden sector $E_8$ group.}, it was realized in the very first paper on string phenomenology \cite{Candelas:1985en}, that the embedding (with natural gauge unification)
\begin{equation}
SU(3) \times SU(2) \times U(1) \subset H \subset E_8 \ ,
\qquad
H = SU(5) \ , SO(10) \ , E_6 \ ;
\end{equation}
is a conducive way of producing the Standard Model from string theory.
To obtain the grand unified theory (GUT) group $H$, one simply has to compactify the $E_8$ heterotic string on a smooth Calabi-Yau threefold with a vector bundle $V$ whose gauge group $G$ is the commutant of $H$ in $E_8$.
Remarkably, pure group theory dictates that $G=SU(n)$ with $n=3,4,5$ precisely does the job.
In fact, the 248 adjoint representation of $E_8$ suffices to incorporate the representations of the fundamental particles:
\[
\mbox{
\begin{tabular}{|l|l|}
\hline
$E_{8}\rightarrow G\times H$ & Breaking Pattern \\  \hline\hline
$\rm{SU}(3)\times E_{6}$ & $248\rightarrow (1,78)\oplus (3,27)\oplus (\overline{
3}
,\overline{27})\oplus (8,1)$ \\  \hline
$\rm{SU}(4)\times\rm{SO}(10)$ &$ 248\rightarrow (1,45)\oplus (4,16)\oplus (\overline{4
},\overline{16})\oplus (6,10)\oplus (15,1)$ \\  \hline
$\rm{SU}(5)\times\rm{SU}(5)$ & $248\rightarrow (1,24)\oplus (5,\overline{10})
\oplus (
\overline{5},10)\oplus (10,5)\oplus (\overline{10},\overline{5})\oplus (24,1)$
 \\ \hline
\end{tabular}
}
\]
Next, we can use a discrete Wilson line to break the GUT group $H$ to the Standard Model group.
The special, and historically the first, case of taking $V=TX$ has come to be known as the ``standard'' embedding.
Thus $SU(n)$ bundles, especially with $n=3,4,5$ will be of particular interest to us; these will give phenomenologically viable candidates.

Next, we must ensure that the four-dimensional effective theory is a well-defined QFT.
A basic requirement is that the Green-Schwarz anomaly be canceled \cite{GSW}:
\begin{equation}
\int_X \tr(R \wedge R - F \wedge F) = 0 \ ,
\end{equation}
where $R$ is the Ricci form on $TX$ and $F$ is the field strength of our bundle.
This is precisely the statement of our condition (2) when we use the explicit expression for the second Chern class.
With the realization of the Ho\v{r}ava-Witten setup \cite{Horava:1996ma} and heterotic M-theory \cite{Lukas:1998tt}, this condition can be generalized to that the
difference $c_2(TX) - c_2(V)$ be an effective curve class. This is because one could allow a non-trivial hidden-sector bundle as well as five-brane classes in the bulk to cancel the anomaly.
In fact, almost all the successful three-generation models beyond the ``standard embedding'' of taking $V = TX$ which was first attempted in \cite{Greene:1986bm} has had the need to cancel anomaly in the presence of these extra classes \cite{Braun:2005nv,Bouchard:2005ag,Candelas:2007ac,Braun:2009qy,Anderson:2012yf}.
In this paper, we will instead focus on the ``extremal'' situation which is more stringently constrained.

%%%%%%%%%%%%%%%%%%%%%%%%%%
\section{The Geography of Extremal Bundles}

With the caveat of the presently unsolved mystery whether smooth Calabi-Yau threefolds exists in finitude, the geography of stable vector bundles on these varieties in fact behave in a much nicer way than naively imagined. Topologically, once we fix the rank and first two Chern classes of the stable holomorphic bundle, there are in fact only finitely many possible numerical values the higher Chern classes could take.  

The mathematical reason for this nice behavior is the well-understood boundedness of various families of stable objects as first demonstrated by the construction of the prototype of the quot scheme. In fact, in principle there are bounds given in terms of the Hilbert polynomials, although in practice finding the precise form of an effective formula is not easy (however see e.g. \cite{KollarMatsu}). For (semi)-stable general rank torsion-free sheaves, boundedness was  proved in \cite{maruyama} \cite{simpson}, by relating the corresponding moduli space to the quot scheme.

As mentioned earlier, there are now several established databases of stable vector bundles and their use in the geography problem may be of help in finding effective bounds. These databases include examples of elliptically fibred Calabi-Yau threefolds, the complete intersections in products of projective spaces, as well as the first few cases ($h^{1,1}$ up to 3) of the vast Kreuzer-Skarke list.
Using these constructions as our 'data', we will proceed to explore the bundle moduli space in a certain `coarse' way. Namely, in looking at the numerical invariants of the full ensemble of stable holomorphic vector bundles, we do not fix the base Calabi-Yau threefolds. 

This is inspired by the similar studies in \cite{Candelas:2007ac,Taylor:2012dr} along the same vein. A physical argument for this approach is simply that all the particular Calabi-Yau varieties are merely solutions to the same universal vacua problem arising from the heterotic string theory. A more dynamical approach to this picture can be motivated from  \cite{Candelas:2007ac,Anderson:2010ty}. The findings in the following sections give us more confidence this strategy is not completely unjustified.

%\todo{Maybe we can prove at outset there will be a finite number of such extremal bundles}

%%%%===============
\subsection{Spectral Cover Bundles on Elliptic Calabi-Yau Threefolds}\label{scbe}

The largest set of explicitly stable $SU(n)$ bundles on $X$ are due to the spectral cover construction \cite{spec}; in this case, $X$ is an elliptically fibred threefold admitting a section $\sigma$.
The base of this fibration is a surface $B$ which can only be one of the following \cite{Morrison:1996na}:
\begin{enumerate}
\item Hirzebruch surfaces $\IF_r$ for $r = 0, 1, \ldots, 12$;
\item Blowups of Hirzebruch surfaces $\widehat{\IF}_r$ for $r=0,1,2,3$;
\item Del Pezzo surfaces $d\IP_r$ for $r = 0,1,\ldots,9$;
\item Enriquez surface $\IE$.
\end{enumerate}
In terms of the Chern classes of the tangent bundle $TB$ of the base, we have
\begin{eqnarray}\label{chernX}
\nn
c_1(TX) &=& 0, \\
\nn
c_2(TX) &=& \pi^*(c_2(TB) + 11 c_1(TB)^2) + 12 \sigma \cdot \pi^*(c_1(TB)) \ ,\\
c_3(TX) &=& -60 c_1(TB)^2 \ ,
\end{eqnarray}
where $\pi : X \rightarrow B$ is the projection map of the elliptic fibration.

An $SU(n)$ bundle $V$ over $X$ is given by the spectral data, consisting of the following two pieces~:
\begin{itemize}
\item The {\em spectral cover} $\mathcal{C}_V$~:  this is an $n$-fold cover of the base and is thus a divisor (surface) in $X$ with degree $n$ over $B$, as an element in $H_4(x; \IZ) \simeq H^2(X,\mathbb{Z})$ it is $[\mathcal{C}_V] =  n\ \sigma +  \pi^* \eta$, where $\sigma$ is the class of the zero section, and $\eta$ is an effective curve class in $H^2(B,\mathbb{Z})$.
In order that $V$ is stable, $\cC$ needs to be irreducible, which follows from the constraints that
(a) the linear system $|\eta|$ is base-point free in $B$ and
(b) $\eta - n c_1(B)$ is an effective curve in $B$.

\item The {\em spectral line bundle} $\mathcal{N}_V$~: this is a line bundle on $\mathcal{C}_V$ with first Chern class
$c_{1}(\cN_V)=n(\frac{1}{2}+\lambda)\sigma+(\frac{1}{2}-\lambda) \pi^{*}\eta+(\frac{1}{2}+n\lambda)\pi^{*}c_{1}(B)$.
The parameter $\lambda$ has to be either integer or half-integer depending on the rank $n$ of the $SU(n)$ structure group~:
\begin{equation}
\lambda =  \left\{
\begin{array}{cl}
	m+1/2 & \text{ if $n$ is odd}, \\
	m & \text{ if $n$ is even},
\end{array} \right.
\end{equation}
where $m\in\mathbb{Z}$.
When $n$ is even, we must also impose $\eta = c_1(B) \textrm{ mod } 2$, 
by which we mean that $\eta$ and $c_1(B)$ differ only by an even element of $H^2(B,\mathbb{Z})$.
\end{itemize}

The Mori cone of effective curves on $X$, in this elliptically fibred case, is spanned by 
\begin{equation}
\sigma \cdot \pi^*(C_i) \ , F \in H_2(X; \IZ) \ ,
\end{equation}
where $F$ is the fibre class and $C_i$ are a basis of effective curves in the base $B$.
The relevant intersections are (curve with surface and surface with surface): 
\begin{equation}
\sigma \cdot F = 1 \ , \quad
\pi^*(C_i) \cdot F = 0 \ , \quad
\pi^*(C_i) \cdot \pi^*(C_j) = \pi^*(C_i \cdot C_j) = C_i\cdot C_j F \ , \quad
\sigma \cdot \sigma = \sigma \pi^*(- c_1(TB)) \ , 
\end{equation}
where for the second intersection one uses the fact that one can always choose a fibre which generically missed a pull-back of a curve in the base;
for the third one uses intersection $C_i \cdot C_j$ in the base, giving a point, which then pulls back to a generic fibre; for the last, one uses adjunction.

The holomorphic $SU(n)$ vector bundle $V$ on $X$ can be extracted from the above data by a  Fourier-Mukai transformation: $(\mathcal{C}_V,\mathcal{N}_V) \,\stackrel{FM}{\longleftrightarrow}\, V$.
The Chern classes of $V$ are given in terms of the spectral data as
\begin{eqnarray}
\nn
   c_1(V) &=& 0 \ , \\
\nn
   c_2(V) &=& \sigma \cdot\pi^*\eta 
              + \pi^*\left( 
              \frac{n}{2} \left(\lambda^2 - \frac 1 4\right) \eta \cdot
                 \left(\eta - nc_1(B)\right) - \frac{n^3 - n}{24} c_1(B)^2
              \right) := \sigma \pi^*\eta + c_F F   \ ,  \\
\label{cVspec}
   c_3(V) &=& 2 \lambda \eta \cdot \left(\eta - nc_1(B) \right) \ .  
\end{eqnarray}
Note that $c_2(V)$, which, being an element of $H^4(X;\IZ)$,  is dual to curve classes, is indeed in the basis of $\sigma \cdot \pi^*(C_i)$ and $F$; the coefficient in front of $F$ is just an integer, which we will call $c_F$, prescribed by intersection on the base $B$. Similarly, $c_3(V)$ is just a number.
Likewise, to be absolutely clear, we can write the Chern classes of $TX$ in \eqref{chernX} as above.
In particular, $c_2(TX) = \sigma \cdot \pi^*(12 c_1(TB)) + (c_2(TB) + 11 c_1(TB)^2) F$.
 
Such stable bundles have been constructed on the various surfaces in \cite{Donagi:2004ia} and \cite{Gabella:2008id}, with a catalogue of millions being found in the latter. In fact, the vast number of examples makes it natural to ask whether they can also be used to explore problems about semi-stable sheaves in positive characteristic \cite{Langer}. 

Now, let us examine the strong condition (2) of \eqref{cherncons}, which  immediately implies that
\begin{equation}
\eta = 12 c_1(TB) \ , \quad
c_{2}(B)
   + \left(11+\frac{n^{3}-n}{24}\right) c_{1}(B)^{2}
   - \frac{n}{2}\left(\lambda^{2}-\frac{1}{4}\right)
   \eta \cdot (\eta - n \ c_1(B)) = 0 \ ,
\end{equation}
whereby placing the constraint
\begin{equation}\label{cBcons}
c_{2}(B)
   + c_{1}(B)^{2} \left(11+\frac{n^{3}-n}{24}
   - \frac{n}{2}(12-n)\left(\lambda^{2}-\frac{1}{4}\right)
\right) = 0
\end{equation}
on the Chern classes of $B$ for some $n \in \IZ_{n \ge 2}$ and appropriate  (half-)integer $\lambda$ .

Now, for the 4 categories of surfaces above, their Chern classes are
\begin{equation}\label{cherntableef}
\begin{array}{|l|c|c|c|}\hline
\mbox{Surface } $B$ & c_1(TB) & c_2(TB) & H^2(B; \IZ) \\ \hline \hline
\IF_{r=0,\ldots,12} & 2S + (r+2)E & 4 & \gen{S,E | E^2 = 0, \ S \cdot E = 1, \ S^2 = -r} \\ \hline
\widehat{\IF}_{r=0,\ldots,3} & 2S + (r+2)F + (r+1)G & 5 & 
\begin{array}{l}
\langle  S,F,G | F^2 = G^2 =-1, \ S\cdot F=G\cdot F =1,  \\
\qquad \qquad \ S\cdot G=0, \ S^2 = -r \rangle   
\end{array}
\\ \hline
d\IP_{r= 0,1,\ldots,9} & 3\ell - \sum\limits_{i=1}^r E_i & 3+r & 
\gen{\ell, E_{i=1,\ldots,r} | \ell^2 =1, \ \ell \cdot E_i = 0, \ E_i \cdot E_j=-\delta_{ij}} 
\\ \hline
\IE & c_1(T\IE)^2 = 0 & 12 & \simeq \IZ^{10} \oplus \IZ_2  \\ \hline
\end{array}
\end{equation}

Immediately, we see that the Enriquez surface is ruled out.

For Hirzebruch surfaces, condition \eqref{cBcons} becomes (interestingly, $r$ drops out of the condition):
\begin{equation}
\frac{1}{3} n \left(144 \lambda^2 (n-12)+(n-36) n+431\right)+92 = 0 \ .
\end{equation}
Recalling that $\lambda$ is half-integer for odd $n$ and integer for even $n$, a simply reduction shows that there are no solutions

Simiarly, for blowups of Hirzebruch surfaces, we have
\begin{equation}
\frac{7}{24} n \left(144 \lambda^2 (n-12)+(n-36) n+431\right)+82 = 0 \ .
\end{equation}
Again, there are no solutions to this in (half-)integers.

Finally, for the $r$-th del Pezzo surface, \eqref{cBcons} becomes
\begin{equation}
-\frac{1}{24} (r-9) \left(n \left(144 \lambda^2 (n-12)+(n-36)
   n+431\right)+264\right)+r+3 = 0 \ .
\end{equation}
Again, we substitute $\lambda = m + \frac12$ for odd $n$ and $\lambda = m$ for even $n$ for some $m \in \IZ$, and then reduce over the integers for $r = 0, \ldots,9$.

We conclude, therefore, that
\begin{proposition}\label{prop-spec}
There are no extremal stable $SU(n)$ bundles from the spectral cover construction on an elliptic Calabi-Yau threefold $X$.
\end{proposition}

Obviously, a sharp result about the existence of elliptic fibration structure on general Calabi-Yau threefolds can substantially expand the applicability of this conclusion, at least for vector bundles obtained by spectral cover constructions. 
Even though there currently exists no lower bound in the rank of the Picard group that guarantees an elliptic fibration, it is worth further consideration\cite{kawamata}.

%%%%%%%%%%%%%%%%%%%%%%%%
\subsubsection{A Generalized Hartshorne-Serre Construction}
%%%%%%%%%%%%%%%%%%%%%%%%

Due to the Proposition above, for the goal of constructing extremal bundles it is desirable to have an alternative construction of stable vector bundles on elliptically fibered  threefolds. In this section, we provide such an alternative and consider the following generalization of the Hartshorne-Serre construction\footnote{For an introduction to the Serre construction, see e.g. \cite{OSS}. The reader is also referred to \cite{Braun:2005nv} for its usage in heterotic model building.}.  Let $C$ be any smooth curve in $X$ with genus $g\ge 1$, and let $\omega_C$ be the canonical sheaf on $C$, then we have a vector bundle $V$ in the following sequence

\begin{eqnarray}\label{hs1}
0\rightarrow H^0(C,\omega_C)\otimes {\cal O}_X \rightarrow V \rightarrow {\cal I}_C \rightarrow 0 \ ,
\end{eqnarray}
where ${\cal I}_C$ is the ideal sheaf on $C$.
The middle term actually gives a locally free sheaf; this follows from $H^1(X,{\cal O}_X)=0$ on a Calabi-Yau threefold.\footnote{In fact this construction is also valid on Fano threefolds where $H^1(X,{\cal O}_X)$ holds as well. For a proof of the locally-freeness of $V$, see  e.g. \cite{mae92,Wu}.}

To discuss stability, we need to use the openness of the stability and that the bundle $V$ is a deformation of the direct sum sheaf $({\cal O}_X^{\oplus h^0(\omega_C)}\oplus {\cal I}_C)$ which is itself semi-stable since it has $c_1=0$. Still further geometric considerations are required to establish stability. We will later in the paper sketch some ideas regarding the stability and leave a full account for a separate publication. 
We now discuss other main features of $V$.

Let us look at how $V$ can be easily made {\bf extremal}. It has Chern classes
\begin{eqnarray}
c_1(V) = c_1(TX) \quad , \quad c_2(V)=[C] \ .
\end{eqnarray} 
Thus, if we have $c_2(TX)=[C]$ then we have automatically an ${\bf extremal}$ bundle.  Here we want to mention that $c_2(TX)$ is not always effective, for elliptical fibrations we calculate this shortly and $c_2(TX)$ is indeed a positive linear combination of effective curves in the base and the elliptic fiber. 

The general situation may be discussed in view of pseudo-effectivity of $c_2(TX)$ since Calabi-Yau threefolds are not uniruled. This, in fact its stronger local version,  follows from the proof of Calabi's conjecture \cite{Yau77,Yau78} and implies that $c_2(X)\cdot H \ge 0$ for a general Calabi-Yau threefold but does not guarantee the class $[c_2(TX)]$ contains an effective curve. So in the general case, (\ref{hs1}) cannot always be applied to construct extremal bundles.

The rank of this extremal bundle depends on $h^0(\omega_C)=h^1({\cal O}_C)=g$,  but unlike the algebraic surface case, the value of $g$ cannot be determined uniquely for a space curve of class $[C]$ in a threefold but we can find a range for it. Using the data in table (\ref{cherntableef}) we can calculate easily that in these cases we have 
\begin{equation}\label{chern2ef}
\begin{array}{|l|c|c|c|c|}\hline
\mbox{Surface } $B$ & c_1(TB)^2 & c_2(TB)+11c_1(TB)^2 & c_2(TB)+c_1(TB)^2& \pi^*(c_1(TB))\cdot \sigma \\ \hline \hline
\IF_{r=0,\ldots,12} & 8 & 92 & 12  & |-K_B|\\ \hline
\widehat{\IF}_{r=0,\ldots,3} & 7 & 82 & 12
& |-K_B| \\ \hline
d\IP_{r= 0,1,\ldots,9} & 9-r & 102-10r  & 12 & |-K_B| \\ \hline
\IE & 0 & 12 & 12  & |-K_B|\\ \hline
\end{array}
\end{equation}
The third column is of course just $\chi({\cal O}_B)$ and its value follows from rationality of the base surfaces $B$, i.e. $p_a=0$. 
We see that except for $d\IP_{9}$ and $\IE$ where $(c_2(TB)+11c_1(TB)^2)=12$, all other cases have large values and especially for $\IF_{r}$ and $\widehat{\IF}_{r}$ the values are constant for all $r$.

From this it follows the dual class of $c_2(X)$
\begin{equation}\label{c2efcurve}
c_2(X)=(c_2(TB)+11c_1(TB)^2) [F]-12\,[\sigma\cdot \sigma]
\end{equation}
In general this corresponds to a curve of large genus, mainly determined by the high degree of $-12[\sigma\cdot \sigma]$ viewed as a curve in the base $B$, but also due to the movable fiber components. 

In fact, as can be computed using the Riemann-Roch formula, the genus of a curve in the class of $\pi^*(c_1(TB))\cdot \sigma=-\sigma\cdot \sigma$ is always one, since it belongs to the linear system $|-K_B| $ and is given by the anti-canonical divisor of $B$. Hence, we have an elliptic curve. 
Let us denote this curve by $C_B$.
To calculate the genus of $12C_B$ we can use again the Riemann-Roch formula for curves in surface $B$
\begin{equation}
2g-2=C\cdot (C+K_B)=11\cdot 12 \, c_1(TB)^2 \ ,
\end{equation}
where $C$ is just $-12K_B$. Thus the genus of $C_B$ is 
\begin{equation}
g(C_B)=1+66c_1(TB)^2
\end{equation}
which takes the values $529$, $463$, $(595-66r)$, and $1$ respectively, for $B=\IF_r, \widehat{\IF}_r, d\IP_r, \IE$. In particular, for $d\IP_{r= 0,1,\ldots,9} $, we find $g=595,529,463,397,331,265,199,133,67,1$. 
Additionally, we need to add a multiple of the elliptic fiber $[F]$ from the formula (\ref{c2efcurve}). 

Even with choices of fibers which do not contribute to the total genus, we can conclude that the total curve of class $[c_2(TB)]$ often has very high genus.
Notice that the genus we find here is the arithmetic genus, while geometric genus can be lower if $C_B$ is singular.

This implies that the rank of $V$ given above in \eqref{hs1} is generally too large for phenomenological applications, although it is certainly locally free and extremal by construction.
In comparsion, the usual Serre construction for a curve $C$ in a threefold gives a rank $2$ sheaf which is not necessarily locally free or even reflexive. 
The exceptions are precisely $d\IP_9$ and $\IE$, both of which have $[c_2(TX)]=[C_B]+12[F]$ with $C_B$ and $F$ both elliptic curves and minimal possible value $g\ge1$ and $rk(V)\ge2$. For these geometries, the rank $2$ case can be shown to be stable but for this particular value of rank, they are generally reflexive sheaves and not bundles.

To construct for example rank $3,4,5$ bundles, we modify the construction of $V$ as follows. For any \footnote{We consider $r\ge 2$ so that at least two independent sections (notice they have zeros when $g\ge 1$) may be picked from $H^0(C, \omega_C)\cong H^1({\cal O}_C)$ to generate $\omega_C$.} $r\le g$,
we consider similarly
\begin{eqnarray}\label{hs2}
0\rightarrow {\cal O}_X^{\oplus r} \rightarrow V \rightarrow {\cal I}_C \rightarrow 0 \ .
\end{eqnarray}
When $r\ge 2$ this gives again a locally free sheaf $V$ for an element of ${\rm Ext}^1( {\cal I}_C ,  {\cal O}_X^{\oplus r})$. When $r=1$ instead, $V$ is rank $2$ and generally only reflexive, and by a result of Serre \cite{OSS} the singular points of $V$ are precisely points on $C$ contained in the zeros of the section of ${\omega}_C$  which spans ${\cal O}_X$ in the construction of $V$.

Notice that although we consider this construction for the elliptically fibered Calabi-Yau threefolds, it is valid on all Calabi-Yau threefolds for $C$ effective. The special feature of the elliptically fibered cases is that the tangent bundle's second Chern class $c_2(TX)$ is Poincar${\rm\acute{e}}$ dual to a curve whose class is effective, and hence can be taken as the curve $C$ in the above construction.  For example, when the threefold has Picard number equal to $1$ say for a quintic threefold (in fact any complete intersection in a single projective space), we can also  construct extremal bundles this way.

%%%%===============
\subsection{Monad Bundles on CICYs}
The next class of bundles are constructed over the so-called CICYs (complete intersection Calabi-Yau threefolds) defined by multi-degree homogeneous polynomials in products of complex projective spaces.
These manifolds, numbering 7890, constituted the first database of Calabi-Yau manifolds to be systematically studied \cite{Candelas:1987kf,Gagnon:1994ek}.
Each manifold is represented by a {\it configuration matrix}
\begin{equation}
\label{cy-config}
X = 
\left[\begin{array}{c|cccc}
\IP^{n_1} & q_{1}^{1} & q_{2}^{1} & \ldots & q_{K}^{1} \\
\IP^{n_2} & q_{1}^{2} & q_{2}^{2} & \ldots & q_{K}^{2} \\
\vdots & \vdots & \vdots & \ddots & \vdots \\
\IP^{n_m} & q_{1}^{m} & q_{2}^{m} & \ldots & q_{K}^{m} \\
\end{array}\right]_{m \times K}
\end{equation}
so that the $j^{\rm th}$ column of this matrix contains the multi-degree of the defining polynomial $p_j$ in $\cA =\IP^{n_1} \times \ldots \times \IP^{n_m}$.
In order that the resulting manifold be Calabi-Yau, the condition
$\sum\limits_{j=1}^K q^{r}_{j} = n_r + 1, \ \forall r=1, \ldots, m$
needs to imposed to guarantee that $c_1(TX)$ vanishes.
The Chern classes of $X$ are:
\begin{eqnarray}
\nn
c_1^r(TX) &=& 0 \ , \\
\label{cX-CICY}
c_2^{rs}(TX) &=& \frac12 \left[ -\delta^{rs}(n_r + 1) + 
  \sum_{j=1}^K q^r_j q^s_j \right] \ , \\
\nn
c_3^{rst}(TX) &=& \frac13 \left[\delta^{rst}(n_r + 1) - 
  \sum_{j=1}^K q^r_j q^s_j q^t_j \right] \ . 
\end{eqnarray}
where we have written the coefficients of the total Chern class
$c = c_1^r J_r + c_2^{rs} J_r J_s + c_3^{rst} J_r J_s J_t$ explicitly, with $J_r$ being the K\"ahler form in $\IP^{n_r}$.
The triple-intersection form $d_{rst} = \int_X J_r \wedge J_s \wedge J_t$ is a totally symmetric tensor on $X$ and the Euler number is simply $\chi(X) = d_{rst} c_3^{rst}$.

Stable bundles over the CICYs are partially classified in \cite{Anderson:2007nc,Anderson:2008uw}\footnote{For the original mathematical construction of monads, see, e.g., \cite{horrocks,barthhulek}, and in the context of $(0,2)$ GLSM description of heterotic compactifications, see \cite{Distler:1987ee}.}.
In particular, on the {\it favourable} CICYs, that is, those with $m = h^{1,1}(X)$ so that the K${\rm \ddot{a}}$hler classes on the ambient space all descend perfectly onto the Calabi-Yau manifold, bundles $V$ of the {\it monad} type were studied. 

These are bundles defined by the short exact sequence
\begin{eqnarray} \label{monad}
\nn &&0 \to V \to B \stackrel{f}{\longrightarrow} C \to 0\ ,
\mbox{ where} \\
B &=& \bigoplus_{i=1}^{r_B} \cO_X({\bf b}_i) \ , \quad
C = \bigoplus_{j=1}^{r_C} \cO_X({\bf c}_j)
\end{eqnarray}
are sums of line bundles with ranks $r_B$ and $r_C$, respectively, so that $V=\ker(f)$.
The parametres ${\bf b}_i = b_i^r$ and  ${\bf c}_j = c_j^r$ are respectively sets of $r_B$ and $r_C$ integer vectors, each of length $m$, indexed by $r$.
To guarantee that $V$ is a bona fide bundle (local freeness), the following constraints
\begin{equation}
c^r_j \ge b^r_j \ , \quad \forall i,j,r, \mbox{ and }
\quad
\forall i,j \mbox{ there exists at least one } r \mbox{ such that }c^r_j>b^r_i
\end{equation}
need to be imposed on these integers.
Furthermore, we see that favourability $m=h^{1,1}(X)$ is needed in order to define these line bundles, as descended from restricting the ambient line bundles of the form $\cO_{\IP^{n_1} \times \ldots \times \IP^{n_m}}(\ell_1, \ldots, \ell_m)$.
The Chern classes of the bundles are
\begin{eqnarray}\label{chernvectors}
\nn {\rm rk}(V) &=& r_B - r_C = n  \ , \\
\nn c_1^r(V) &=& \sum_{i=1}^{r_B} b^r_i - \sum_{j=1}^{r_C} c^r_j  = 0\ ,
\\
c_{2}(V)_r &=& \frac12  d_{rst} 
   \left(\sum_{j=1}^{r_C} c^s_j c^t_j- 
   \sum_{i=1}^{r_B} b^s_i b^t_i \right) \ , 
\label{chernVmonad} \\
\nn c_3(V) &=& \frac13 d_{rst} 
   \left(\sum_{i=1}^{r_B} b^r_i b^s_i b^t_i - \sum_{j=1}^{r_C} c^r_j
   c^s_j c^t_j \right) \ ,
\end{eqnarray}
where $d_{rst}$ are the triple intersection numbers on $X$.
For the second Chern class, we have lowered the double index into a vector by contraction with $d_{rst}$, i.e., $c_{2}(V)_r = d_{rst} c_2(V)^{st}$.

The constraint (2) of \eqref{cherncons}, here translating to $c_2(V) = c_2(TX)$  since $c_1(TX)=c_1(V)=0$ , thus implies that, for all $s,t = 1, \ldots, m$,
\begin{equation}\label{c2CICY}
\sum_{j=1}^{r_C} c^s_j c^t_j - \sum_{i=1}^{r_B} b^s_i b^t_i  =
\sum_{j=1}^K q^s_j q^t_j - \delta^{st} \sum\limits_{j=1}^K q^{s}_{j} \ .
\end{equation}
The algorithmic approach allows us to quickly scan through the available datasets, in order to analyze this equation in detail.

%%%%
\subsubsection{Positive Monads}\label{posmon}

For so-called {\it positive monads} where the entries $b^r_i$ and $c^r_j$ are all positive integers, it was proven in \cite{Anderson:2008uw} that there are a finite number of these bundles on the favourable CICYs and that the majority of them a stable.
The dataset, motivated by physics, was compiled for $SU(n)$ with $n=3,4,5$,
totaling 7118 bundles on the 4515 manifolds.
The positive requirement on the degree of the line bundles is very severe, and results in the small number of available choices up to rank $\le 5$.

A scan over these shows that there are actually quite a number of bundles which satisfy our extremality conditions:
respectively 305 over 20 manifolds, 729 over 12 manifolds and 25 over 7 manifolds for $n=3,4,5$.
We can see the distribution of $\ch_3(V)$ for each of the cases:
\[
\begin{array}{ccc}
\includegraphics[trim=0mm 0mm 0mm 0mm, clip, width=2.0in]{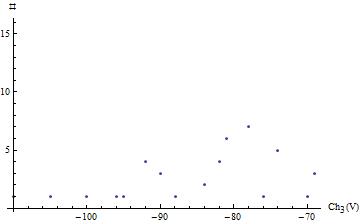}
&
\includegraphics[trim=0mm 0mm 0mm 0mm, clip, width=2.0in]{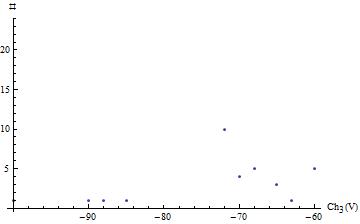}
&
\includegraphics[trim=0mm 0mm 0mm 0mm, clip, width=2.0in]{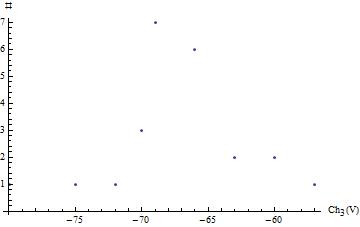}
\\
n=3 & n=4 & n=5
\end{array}
\]

\paragraph{Three-Generation Candidates: }
For reference, let us present the ones which are potential three-generation candidates.
That is, those which also satisfy a third condition
\begin{equation}\label{3gen}
(3) \qquad \ch_3(V) = 3k \ , \mbox{ with } k \in \IZ \mbox{ dividing } \chi(X) \ .
\end{equation}
This is so that if we were to find an appropriate freely acting discrete group $G$ of order $k$ on $X$, then on the quotient manifold $X/G$, there would be exactly $\ch_3(V/G) = 3$ generations in heterotic compactification.
Obviously $k$ must divide the Euler number $\chi(X)$ in order for $G$ to be freely acting.

There is only a total of 13 of bundles satisfying conditions (1) - (3).
These are:
\begin{itemize}
\item $SU(3)$ Bundles:
There are 5, all living on the bi-cubic manifold 
$\left[\begin{array}{c|cc}
\IP^{2} & 3 \\
\IP^{2} & 3 
\end{array}\right]$, which has $c_2(TX) = (36,36)$ and $\chi(TX) = -162$.
We list these below as
\begin{equation}
\begin{array}{|c|c|c|c|}\hline
B = \bigoplus\limits_{i=1}^{r_B} \cO_X({\bf b}_i) & 
C = \bigoplus\limits_{j=1}^{r_C} \cO_X({\bf c}_j) &
\ch_3(V) = 3k &  \chi(X) / k \\  \hline \hline
\cO(1,2) \oplus \cO(1,1)^4 & \cO(4,2) \oplus \cO(1,4) & -81 & 6 \\ \hline
\cO(2,1) \oplus \cO(1,1)^4 & \cO(4,1) \oplus \cO(2,4) & -81 & 6 \\ \hline
\cO(1,2)^4 \oplus \cO(1,1)^3 & \cO(4,2) \oplus \cO(1,3)^3 & -81 & 6 \\ \hline
\cO(2,1)^4 \oplus \cO(1,1)^3 & \cO(2,4) \oplus \cO(3,1)^3 & -81 & 6 \\ \hline
\cO(2,1)^3 \oplus \cO(1,1)^3 \oplus \cO(1,2)^3 
& \cO(2,2)^6 & -81 & 6 \\
\hline
\end{array}
\end{equation}

\item $SU(4)$ Bundles:
There are 7 in total.
On the intersection of two cubics in $\IP^5$, i.e.,
$X = [\IP^5 | 3, 3]$, which has $c_2(TX)=54$ and $\chi(X) = -144$, there is one bundle
\begin{equation}
B = \cO(1)^6 \ , \quad
C = \cO(3)^2 \ ,
\end{equation}
whose $\ch_3(V) = -72$.

On $X=[\IP^6 | 3,2,2 ]$, which has $c_2(TX)=60$ and $\chi(X)=-144$, there is one bundle
\begin{equation}
B = \cO(1)^7 \ , \quad
C = \cO(3) \oplus \cO(2)^2 \ ,
\end{equation}
whose $\ch_3(V) = -72$.

On $X = \left[\begin{array}{c|cc}
\IP^{2} & 2 \\
\IP^{2} & 4 
\end{array} \right]$, with $c_2(TX) = (24,44)$ and $\chi(X) = -168$, there are 5:
\begin{equation}
\begin{array}{|c|c|c|c|}\hline
B = \bigoplus\limits_{i=1}^{r_B} \cO_X({\bf b}_i) & 
C = \bigoplus\limits_{j=1}^{r_C} \cO_X({\bf c}_j) &
\ch_3(V) = 3k &  \chi(X) / k \\  \hline \hline
\cO(1,2) \oplus \cO(1,1)^6 & \cO(5,2) \oplus \cO(1,3)^2 & -72 & 7 \\ \hline
\cO(1,2)^2 \oplus \cO(1,1)^6 & \cO(3,2)^2 \oplus \cO(1,3)^2 & -72 & 7 \\ \hline
\cO(1,2)^2 \oplus \cO(1,1)^6 & \cO(4,2) \oplus \cO(2,2) \oplus \cO(1,3)^2 & -72 & 7 \\ \hline
\cO(1,2)^3 \oplus \cO(1,1)^6 & \cO(3,2) \oplus \cO(2,2)^2 \oplus \cO(1,2)^2 & -72 & 7 \\ \hline
\cO(1,2)^4 \oplus \cO(1,1)^6
& \cO(2,2)^4 \oplus \cO(1,3)^2 & -72 & 7 \\
\hline
\end{array}
\end{equation}

\item $SU(5)$ Bundles:
There is only one on the quintic $[\IP^4 | 5]$, with $c_2(TX) = 50$ and $\chi(X)=-200$, with
\begin{equation}
B = \cO(2)^5 \oplus \cO(1)^5 \ , \quad
C = \cO(3)^5 \ ,
\end{equation}
with $\ch_3(V) = -75$.
\end{itemize}

We see that though these bundles exactly cancel anomaly and are potential candidates for three generations, there are none which have precisely $\ch(TX)=\pm 3$, or $k = 1$.
\comment{
Now, if the bi-cubic had a freely-acting order 27 group, on the quotient we would have the perfect solution.
There is a standard $\IZ_3 \times \IZ_3$ freely acting group; however, I am not aware of anything beyond this.
I hope that maybe the covering group, which is a $\IZ_3$ lift of this group, would exist in special points in moduli space.
}
The particularly interesting model is that last one.
It is well known that the quintic $Q=[\IP^4|5]$ has a freely acting $\IZ_5 \times \IZ_5$ discrete symmetry; our $SU(5)$ bundle $V$ defined by
\begin{equation}
0 \to V \to \cO_Q(2)^5 \oplus \cO_Q(1)^5 \to  \cO(3)^5 \to 0
\end{equation}
is actually equivariant with respective to this discrete group.
This was found in \cite{Anderson:2007nc} and then discussed in nice detail in \cite{Braun:2009mb}.
Therefore, on the quotient manifold $Q' \simeq Q / \IZ_5 \times \IZ_5$, $V$ descends to an $SU(5)$ bundle $V'$ with $\ch_3(V') = -3$ and $\ch_2(V) = \ch_2(Q')$.

%%%%===============
\subsection{Poly-stable Bundles}
The last large class of bundles we wish to consider are perhaps the simplest of them all.
These are direct sums of line-bundles.
Now, a line-bundle is by definition stable since there are no injective sub-sheaves which could de-stablize it.
However, in the case of a direct sum of line bundles, we need to be careful.

What we are after are solutions of the Hermitian Yang-Mills equations and the Donaldson-Uhlenbeck-Yau Theorem establishes existence of such solutions to {\it poly-stable}  vector bundles, which are direct sum over stable bundles of the {\it same slope}.
In fact, on each `S-equivalent' orbit there exists such polystable bundles as semi-stable limit points. 
In \cite{Anderson:2012yf,Anderson:2013xka}, this was exploited to construct Standard-Model vacua and has given the largest database of such models.

Let us, therefore, attempt to find such simple poly-stable bundles on the CICYs.
That is, we look for
\begin{equation}
V = \bigoplus\limits_{i=1}^n L_i := 
\bigoplus\limits_{i=1}^n\cO_X(\ell_i^1, \ldots, \ell_i^m) \ ,
\qquad \sum\limits_{i=1}^n \ell_i^r = 0 \ , \forall \ r = 1, \ldots, m \ ,
\end{equation}
where $m$ is, as before, the number of factors in the product of projective spaces and the sum condition on the integer parametres $\ell_i^r$ ensures that we have an $S(U(1)^n)$ bundle.
Again, we will focus on {\it favourable} CICYs and the definition of the {\it slope} $\mu(L)$ for a line bundle $L = \cO_X(\ell^1, \ldots, \ell^m)$ is
\begin{equation}
\mu(\cO_X(\ell^1, \ldots, \ell^m)) := \int_X c_1(L)\wedge J^2 = 
d_{rst} \ell^r T^s T^t \ , \qquad T^{r = 1, \ldots, m} \in \IZ_{\ge 0} \ ,
\end{equation}
for some polarization $J := \sum\limits_{r=1}^m T^r J_r$ in the K\"ahler cone.
Subsequently the polystability condition becomes
\begin{equation}
\mu(L_i) = \mu(V) = 0 \ 
\leadsto
\exists T^r > 0 \mbox{ such that } \ d_{rst} \ell_i^r T^s T^t = 0 \ ,
\quad \forall i = 1,\ldots,n \ .
\end{equation}
Of course, we want non-trivial solutions so that not all $T^r=0$.
In fact, we have imposed, in the above, the more stringent condition that we are in the interior of the K\"ahler cone so that $T^r > 0$.
Immediately, we see that the integer entries $\ell_i^r$ must be a mixture of positive and negative integers.

From \eqref{cX-CICY} and by setting the $C$ to 0 in \eqref{chernVmonad} to quickly obtain the Chern classes of direct sums of line bundles, we see that the extremality condition (2) of \eqref{cherncons} translates to
\begin{equation}
-\delta^{rs}(n_r + 1) + 
  \sum_{j=1}^K q^r_j q^s_j
=-  \sum_{i=1}^{n} \ell^r_i \ell^s_i \quad \forall \ r,s=1,\ldots,m \ .
\end{equation}
Searching through the publicly available database of the 202 Standard-Model bundles from \cite{Anderson:2012yf}, we see that {\bf none} is actually extremal.

%%%===============
\subsection{A Side on Donaldson-Thomas Invariants}
Of course, we have only touched upon the information hidden in the positive monad dataset as an illustration.
Relaxing the entries to have non-positive integers is perfectly allowed though we loose control over finiteness in the classification.
Moreover, it is more difficult to prove stability for such a general case except when Picard number is one.
Therefore,  one way of enumeration is by simply attacking the Diophatine system given in \eqref{c2CICY} which only contain topological information about the vector bundles.  A more complete description of  the monad bundles will require computing deformation moduli. This will give us the local geometry of the holomorphic deformations which 
are also captured by the holomorphic Chern-Simons functional. One may ask whether one can evaluate DT invariants for the model building motivated examples of stable vector bundles.

Though explicit expressions for certain classes of DT invariants have been obtained in the literature following the localization methods of \cite{MOOP}, and there have been some nice developments, comprising of interesting combinatorics and functional equations satisfied by these invariants \cite{KS,BBS,JS}\footnote{Note that in particular in \cite{JS}, the identification of the local geometry in deformation space for general stable coherent sheaves with that of the neighborhood of critical points of the holomorphic Chern-Simons functional permits one to generalize the definition of Donaldson-Thomas invariants. The holomorphic connection on a vector bundle which comprise these critical points is precisely what we are interested in within the heterotic string context, namely Hermitian Yang-Mills solutions on Calabi-Yau threefolds. }, the cases of particular interest for 
heterotic model bundle constructions remain less explored, not least due to their higher rank. In addition, the examples studied in the literature are mostly sheaves which do contain `anomaly' and will not be extremal. 
For example, results have been obtained for certain DT invariants \cite{LQ} on the CICY $[\IP^5 | 2,4]$. Unlike $c_1(V)$ which can be easily changed for example by twisting, changing $c_2(V)$ to satisfy anomaly cancellation requires more subtle local modifications of the 
bundle, which can make it no longer locally-free or even torsion. In this sense, even finding out the asymptotics of DT invariants for the Heterotic bundle constructions are very nontrivial tasks. Since our main interest in the present paper is the numerical invariants rather than bundle moduli, we will not try to discuss this further.

%%%%%%%%%%%%%%%--------------------------------------------
%%%%==============
\section{The DRY Conjecture}\label{DRYconj}

We now turn to general results on possible numerical values realized by Chern classes of stable vector bundles on Calabi-Yau manifolds. It is perhaps worth stressing that most of our examples have natural embeddings in (weighted) projective space and this warrants our treatment of the bundle Chern classes simply as numbers, a more intrinsic point of view is to treat these classes and their numerical relations as reflections of positivity for the vector bundles. 

Both from the mathematics and physics points of view, it is very difficult to predict what values of Chern classes are populated by actual vector bundles, and of course much more difficult to answer the same question with the further cndition of stability. 
Subsequently, it is extremely interesting to have existence criterions for bundles based on Chern class values. One conjecture of this type is due to Douglas-Reinbacher-Yau who proposed the following  \cite{Douglas:2006jp} \footnote{See also subsequent work in \cite{Andreas:2010rc,Andreas:2011zs}, which focuses on bundles with vanishing first Chern class, as is in our case.}

\begin{conjecture} [DRY]
On a simply-connected Calabi-Yau threefold, a stable vector bundle $V$ of rank $n$ and $c_1(V)=0$ with prescribed $c_2(V)$ and $c_3(V)$ will exist if
\[
c_2(V) = n (H^2 + \frac{1}{24} c_2(TX)) \ , \qquad
c_3(V) < \frac{16\sqrt{2}}{3}\ n \ H^3 \ ,
\]
for some ample class $H \in H^2(X, \IR)$.
\end{conjecture}
More exactly, it should be noted that the proposal is actually concerned with the more general class of sheaves called reflexive sheaves instead of vector bundles. Results about bounds on Chern numbers of low rank reflexive sheaves on especially $\IP^3$ are known in the literature, but they usually follow from application of Riemann-Roch and hence says nothing about existence, and are much less tight, in the sense that $c_3(V)$ is only bounded by $[c_2(V)]^2$ in numerical value.

For the extremal bundles considered in this paper, we can impose additionally $c_2(V) = c_2(TX)$, and the DRY conjecture simplifies to
\begin{equation}\label{duy}
\frac{24 - n}{24 n} c_2(TX) = H^2 \ , \qquad
\frac{3}{16n\sqrt{2}} c_3(V) < H^3 \ .
\end{equation}
Note that the first equation coming from extremality fixes the polarization $H$ and we only need to check the inequality part of the conjecture. 
Due to ampleness of $H$ it can be inferred that $n<24$ when $c_2(TX)>0$ (more precisely nef).

We will now find some explicit solutions to these equations and inequalities.
Incidentally, this constraint seems to point to some kind of relevance of $K3$ regarding existence of stable bundles on Calabi-Yau threefolds. Although it is not known whether each Calabi-Yau threefold contains a $K3$, one may certainly consider  a stable degeneration into a pair of Fano threefolds  glued along a $K3$ surface.

%%%================
\subsection{Explicit Solutions to DRY}\label{exDRY}

Let us now impose the multiple-of-three-generation condition \eqref{3gen} as a quick way to reduce the difficulty of our numerical task. Then $c_3(V)$ is purely numerical and we can find explicit solutions
to the DRY conditions by parametrizing the ample class $H$ when the CICY involved is favourable.
It is easy to see that upon this constraint, the DRY conjecture becomes a purely topological condition on the base Calabi-Yau threefold:
\begin{equation}
\frac{24 - n}{24 n} c_2(TX) = H^2 \ , \qquad
\frac{9k}{8n\sqrt{2}} < H^3 \ , k \in \IZ, k \mbox{ divides } \chi(X) \ .
\end{equation}
This allows us to explicitly check the DRY conjecture in \eqref{duy} on the CICYs to see how many manifolds satisfy the constraints.
Here, the ample class can be written simply as $H = \sum\limits_{r=1}^{h^{1,1}(X)} h_r J^r$ with $h_i \in \IR_{\ge 0}$.
Note that to be general we need not restrict our attention to only the favourable ones.
Using the intersection form $d_{rst}$ and topological data in \eqref{cX-CICY}, we can write \eqref{duy} explicitly as
\begin{align}
\nn
?\exists h^s \in \IR_{\ge 0} \ , k \in \IZ  \quad : \qquad & 
\frac{24-n}{48n} d_{rst} \left[ -\delta^{rs}(n_r + 1) + 
  \sum_{j=1}^K q^r_j q^s_j \right] = d_{rst} h^s h^t \ , \\
\nn
&\frac{9k}{8n\sqrt{2}} < d_{rst} h^r_j h^s_j h^t_j  \ , \\
&\chi(X)/k \in \IZ
\ ; 
\end{align}
with implicit summation on repeated indices $r,s,t = 1, \ldots, h^{1,1}(X)$.

We scan through all 7890 CICY manifolds, and find solutions for $n=3,4,5$ for all the manifolds, except 37 manifolds.
Curiously, these are all with $\chi(X) = 0$, including trivial cases such as $T^2$ or $T^2 \times K3$.
Interestingly, the famous Schoen self-mirror manifold with Hodge numbers $(19,19)$, which is a CICY with configuration $X = \left[\begin{array}{c|ccc}
\IP^{2} & 3 & 0\\
\IP^{2} & 0 & 3\\ 
\IP^{1} & 1 & 1\\ 
\end{array}\right]$, is among these exceptions.

Without much difficulty, one would imagine that these manifolds admit other constructions of stable vector bundles than monads, nonetheless their uniformly being self-mirror with respect to the Hodge diamond is amusing.

%%%%%%%%%%%%%%%%%%%
%\subsection{Poly-stable Bundles and DRY}
%\todo{I have access only to about 200 of these bundles, so maybe we can skip this section? Or I can initiate a new scan}

%%%%%%%%%%%%%%%%%%%%%%%
\subsection{Bolgomolov Bound }\label{bbef}

In the presence of an elliptic fibration structure (for existence of such structures see \cite{oguiso,wilson:e}), it makes sense to examine the Bogomolov bound in view of restriction theorems for stability. 
We recall that a bundle $V$ of rank $n$ slope-stable with respect to a K\"ahler class polarization $J$ on the threefold $X$ must (necessarily but not sufficiently) satisfy the Bogomolov bound that \cite{bogomolov}:
\begin{equation}
\int_X (2 n c_2(V) - (n-1) c_1(V)^2 )\wedge J > 0 \ .
\end{equation}
For our unitary bundles and for our polarization $H$, this simply translates to
$c_2(V) \wedge H > 0$.
The most general ample polarization class $H$, being in $H^2(X;\IZ)$, can be expressed in terms of divisor (surface) classes and thus can be written as \cite{Andreas:2010rc}:
\begin{equation}\label{polarH}
H = s \sigma + \pi^*(\rho) \ , \quad
s \in \IZ_{>0} \ , \quad
\rho - s \ c_1(TB) \mbox{ ample on } B \ .
\end{equation}
Checking ampleness on $B$ is relatively straight-forward: one simply has to make sure that the curve intersected with the generators of the Mori cone on $B$ gives positive integers.
Indeed, $V$ is stable for spectral cover bundles \cite{Friedman:1997ih} if $H = H_0 + h' \pi^*(H')$ for some generic $H_0 \in H^2(X;\IZ)$ and $H' \in H^2(B;\IZ)$ but for sufficiently large $h' \in \IZ_{\gg 0}$.
How sufficiently large seems not to be known in general.

Similarly, for $V$ to be stable, the spectral cover $\cC$ needs to be effective and irreducible.
It suffices, therefore, to require that it be ample.
Let us first stick to this sufficient but not necessary condition, and then attempt to relax it.
In the same way as $H$, the ampleness of $\cC$ implies that
\begin{equation}
\cC = n \sigma + \pi^*(\eta) \ , \qquad
\eta - n \ c_1(TB) \mbox{ ample on } B \ .
\end{equation}

Expand the intersection using \eqref{cVspec}, we have the Bogomolov term being 
\begin{equation}\label{c2VH}
c_2(V) \cdot H = -s (c_1(TB) \cdot \eta) + \eta \cdot \rho + s c_F \ .
\end{equation}
Recalling the expression for $c_F$ from \eqref{cVspec}, we see that the term $\eta \cdot (\eta - n c_1(TB))$ is always positive for ample $\cC$ because $\eta - n c_1(TB)$ was required to be ample, hence for odd $n$, the miminum of of the positive term in $c_F$ is attained for $\lambda = \pm \frac12$, and for even $n$, $\lambda = \pm 1$.

We thus have that
\begin{equation}\label{theconstraints}
c_2(V) \cdot H \ge
\left\{
\begin{array}{lcl} 
\left[ \eta \cdot (\rho - s c_1(TB)) \right]
- \frac{n^3 - n}{24} s c_1(B)^2
& & n \mbox { odd } \\
\left[
\eta \cdot (\rho - s c_1(TB)) + 
\frac{3ns}{8} \eta \cdot \left(\eta - nc_1(B)\right) 
\right] - \frac{n^3 - n}{24} s c_1(B)^2
& & n \mbox { even . } \\
\end{array}
\right.
\end{equation}
Again, due to the ampleness of $\rho - s c_1(TB)$ %as is required by ampleness of the polarization, 
the first term in the square brackets is always positive.

For concreteness, consider a particular threefold, say the elliptic fibration over the $r$-th Hirzebruch surface.
Recall that the Mori cone is the positive span of $S$ and $E$ such that $E^2 = 0, \ S \cdot E = 1, \ S^2 = -r$, together with $c_1(TB) = 2S + (r+2)E$ and $c_2(TB) = 4$.
On this manifold, the spectral curve $\eta = a S + b E$ for $a,b, \in \IZ_{\ge 0}$ and the ample curve $\rho = h_1 S + h_2 E$, so that constraints (\ref{theconstraints}) become
\begin{align}
\nn
s \ , \quad h_1 - 2s \ , \quad h_2 - h_1 r + (-2 + r) s > 0 \ ,
\\
a > 2n \ , \quad b > n(r+2) \ , \quad b \ge r a \ .
\end{align}
Now, let us find $h_1$ and $h_2$ which minimally satisfy these constraints.
Therefore, we can set $s=1$, $h_1 = 2s+1 = 3$, $a = 2n+1$, 
$b = \max[n(r+2)+1, r a]$, whence, $c_2(V) \cdot H \ge$
\begin{equation}
\left\{
\begin{array}{lcl} 
\max [2 n r+r,n (r+2)+1]+2 h_2 n+h_2-\frac{1}{3} n
   \left(n^2+12 r+11\right)-2 (r+1) &&  n \mbox { odd } \\
\frac{1}{24} (6 (3 n (n+1)+4) \max [2 n r+r,n (r+2)+1]+&&\\
\qquad 24 h_2 (2
   n+1)-3 (3 n (n+1)+16) (2 n+1) r-2 (n (n (22 n+9)+44)+24))
 &&  n \mbox { even } \\
\end{array}
\right.
\end{equation}
Therefore, it suffices to take, minimally, $h_2$ to be $h_1 r + (2-r)s = 2r+2$ plus some $n$-dependet positive correction. Specifically, we can take
\begin{equation}
h_2 = \left\{
\begin{array}{lcl} 
2r+2 + \frac16 h^2
& & n \mbox { odd } \\
\frac{1}{24} n (n (9 r+22)+9 r-2)
& & n \mbox { even .} \\
\end{array}
\right.
\end{equation}

%%%%%%%%%%%%%%%%
\subsubsection{Bogomolov versus DRY}\label{bbDRY}

From the DRY Conjecture, we have the following inequalities:
\begin{equation}\label{Fr-DRYBogo}
c_2(V) = n(H^2 + \frac{1}{24} c_2(TX)) \ , \quad
c_2(V)H > 0 \ , \quad
c_3(V) < \frac{16 \sqrt{2}}{3} n H^3 \ .
\end{equation}
Continuing the notation for the polarization in \eqref{polarH}, we have that
\begin{align}
\nn
H^2 &= s \sigma^2 + \pi^*(\rho^2) + 2s \sigma \pi^*(\rho) 
 = s \ \sigma \pi^*(2 \rho -c_1(TB)) + \rho^2 \ F \ , \\
H^3 &= s^2 (c_1(TB)^2 - 2 \rho \cdot c_1(TB)) + s (2 \rho^2 - \rho \cdot c_1(TB)) + s \rho^2 \ .
\end{align}
In fact, DRY is strong enough to fix the polarization $H$ by the equality in $c_2(V)$, since $c_2(V) - \frac{n}{24} c_2(TX) = n H^2$, we have that
\begin{align}\label{Bogo-DRY}
\nn
\eta - \frac{n}{2} c_1(TB) & = s \ n \ (2 \rho - c_1(TB)) \ , \\
c_F - \frac{n}{24} ( c_2(TB) - 11 c_1(TB)^2 ) & = n \rho^2 \ .
\end{align}

Again, use the same example as above, namely the r-th Hirzebruch surface as base for the elliptic fibered threefold.
We have that $\rho$ and $\eta$ as above, so that \eqref{Bogo-DRY} becomes
\begin{equation}
(a-n - (2 h_1  -2) s n) S + (b - \frac{n(r+2)}{2} - (2 h_2 - (r+2))sn) E = 0
\ , \quad
c_F + \frac{7}{2} n = n h_1 (2 h_2 - h_1 r) \ .
\end{equation}
This is just enough to solve for the polarization: three equations in $(s,h_1,h_2)$; the system can be solved exactly though the expression is too long to present here.
We can substitute back the solution into the remaining inequalities and study the region in $(a,b,\lambda)$ space.

Although we can find solutions to Bogomolov inequality in the considered Calabi-Yau threefolds with elliptic fibration, for restriction theorem to imply stability on the threefold, the restrictions must be to a general divisor. We have also only considered Bogomolov inequality associated to the zero sections of elliptic fibration which is clearly special. 

Very generally, establishing Bogomolov inequalities for particular divisors is very far from establishing the same result for a `general' divisor. So by restricting and comparing Bogomolov with DRY we can mostly just obtain consistency checks, instead of establishing a direct relation between the two. Nevertheless, our examples seem to suggest that to a large degree, Chern numbers consistent with the DRY conjecture are actually populated by vector bundles.

%%%%%%%%%%%%%%%%%%%%%%
\subsection{More on the Range of $c_3(V)$}\label{c3range}

In this section, we give some ideas about stability of the bundle constructed by the extensions (\ref{hs1}) and (\ref{hs2}). That its stability has implications for DRY conjecture will be demonstrated. First we observe the simple relation between $c_3(V)$ and $g_{C}$. On the one hand, from the short exact sequence of ideal sheaves
\begin{equation}
0\rightarrow {\cal I}_C \rightarrow {\cal O}_X \rightarrow {\cal O}_C \rightarrow 0 \ ,
\end{equation}
we have the formula
\begin{equation}
\chi({\cal I}_C)=-\chi({\cal O}_C)=g-1 \ ,
\end{equation}
which follows from $\chi({\cal O}_X)=0$ for Calabi-Yau threefold. 

On the other hand, from the construction of $V$,
\begin{equation}
ch_i(V)=ch_i({\cal I}_C) \ .
\end{equation}
Applying Riemann-Roch using Chern classes of $X$ given in  equation (\ref{chernX}) and noticing $c_1(V)=0$ we find simply
\begin{equation}
\chi(V)={1\over2}c_3(V) \ .
\end{equation}
%\begin{equation}
%\chi(V)%=\int_X \, ch({V})Td(TX)
%={r_Vc_1(X)c_2(X)\over 24}+{c_1(X)^2+c_2(X)\over 12}c_1(V)+{c_1(V)^2-2c_2(V)\over 4}c_1(X)+ch_3(V)
%\end{equation}
Comparing these two results we conclude that 
\begin{equation}
c_3(V)=2g_C-2 \ .
\end{equation}
So once we have picked a curve $C$ in class $[C]$, we can determine the third Chern class of $V$. Hence in this construction, $c_2(V)$ and $c_3(V)$ are tied together by geometry. 

While the arithmetic genus $g_C$ remains invariant when we deform the complex structure of $V$, as mentioned above we may choose different curves $C$ in the class of $[C]$ with different values of $g_C$.
In the spirit of the DRY conjecture, we need to establish a bound on the value of $g_C$ given the class of $[C]$.
Of course, determining the genus of an algebraic curve in a threefold is very nontrivial, and we shall sketch only some rough points in this paper.

A common feature for all of our elliptic fibered Calabi-Yau threefolds is that the normal bundle of any curve in $B$ is negative in the fiber direction and so such curves cannot be moved out of $B$. This comes from the fact that $N_{B\backslash X}=K_B$ in order that $c_1(TX)=0$. 
However if we combine with the right number of elliptic fibers $F$, the new contributions can give the normal bundle sections.

Concretely, let us consider the simplest example with $B=\IP^2$, and take a degree $1$ divisor in the base,  then we may identify it with $H$ the hyperplane class. This is simply a line, and the discriminant divisor of the elliptic fibration intersects with it with multiplicity $36$. Thus we find an elliptically ruled surface with $36$ singular fibers.  

This $\IP^1$ has normal bundle ${\cal O}(-3)\oplus {\cal O}(1)$, so we may consider any integer $m\ge 2$ and 
take curves in the classes $m(H+aF)$ with $a\ge 3$, whose normal bundle is now nonnegative. Since curves are unobstructed in the surface we can move it to another curve in the same linear system. 
For the $B=\IP^2$ example, we may consider moving for example the component $34(H+3F)=34H+102F=C-2H$ of the curve associated to $c_2(TX)$.

During such a deformation, the curve has constant arithmetic genus, so we have not changed the value of $c_2(V)$ in our construction. However, once the curve has been moved out of the base $B$, we can find directions in which local modifications can be made to lower its genus, while preserving its curve class (lower dimensional schemes may be attached). A bound on the amount of change in $g_C$ will then lead to a relation between $c_2(V)$ and $c_3(V)$ which can be compared against the DRY conjecture.
A detailed account of this procedure will be published elsewhere\footnote{Although we limited ourselves here to the curve $C$ in class of $c_2(TX)$ in the Serre construction for the goal of getting extremal bundles, this is certainly unnecessary and would be removed in studying DRY conjecture. Our considerations apply just as well to any effective curve $C$.}.

%%%%%%%%%%%%%%%%%%%%
\section{Statistics of Chern Classes}\label{chernstat}

In this last part, we include some simple  statistical analysis of  Chern classes of stable $SU(n)$ bundles on Calabi-Yau threefolds.
In particular, we can plot $\ch_3(V)$ versus the {\it difference} between $\ch_2(TX)$ and $\ch_2(V)$.
We will define this difference as needed in the cases below. In some sense, there is no natural `pairing' or `norm' for these classes which belong to $H^4(X,\IZ)$ on a threefold, unlike cubic forms. Nevertheless, the vector approach we shall take is well justified, especially as it includes generators of the nef cone. 

From physics considerations, these plots are also very natural. They give a `correlation' of the Green-Schwarz anomaly carried by the stable vector bundles to that of the generation number of the underlying Calabi-Yau threefold. 

We find quite intriguing patterns in the case of elliptically fibered Calabi-Yau threefolds in these distributions. Their contours suggest that there is indeed quite universally shaped bounds on these Chern numbers' numerical values. They also appear very sensitive to parity of the rank of the bundle, i.e. the value of $r$ mod $2$. Here we also observe that power of these database of examples, in comparison, the `positive' monad example lacks in abundance and as a result, not much can be said about their distribution at present. 

%%%
\subsection{Stable Monads on CICYs}
\begin{figure}[!h!t!b]
\centerline{
(a)
\includegraphics[trim=0mm 0mm 0mm 0mm, clip, width=2.0in]{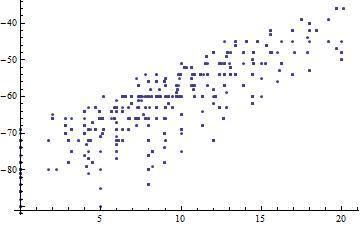}
(b)
\includegraphics[trim=0mm 0mm 0mm 0mm, clip, width=2.0in]{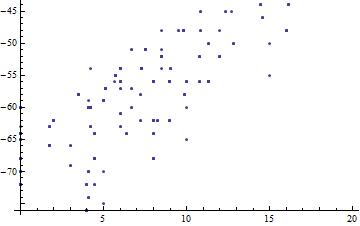}
(c)
\includegraphics[trim=0mm 0mm 0mm 0mm, clip, width=2.0in]{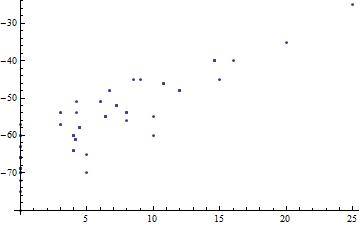}
}
\caption{{\sf {\small
A plot of $\ch_3(V)$ in the ordinate versus $| \ch_2(TX)_r - \ch_2(V)_r |$ in the abscissa for the 7118 $SU(n)$ positive monads on CICY manifolds for $n=3,4,5$, shown respectively in (a), (b) and (c).
}}
\label{f:posmonads}}
\end{figure}

A good starting point is the space of positive monad bundles on the favourable CICYs.
We recall that here we have the second Chern classes being defined as a vector (see (\ref{chernvectors})), one natural difference we can define is simply the norm of the difference between the two vectors, i.e.,
$| \ch_2(TX)_r - \ch_2(V)_r |$.
We therefore plot this difference versus $\ch_3(V)$ for $n=3,4,5$ for the 7118 positive bundles in Figure \ref{f:posmonads}.

%%%
\subsection{Extremal Spectral Cover Bundles}\label{scebstat}
In a similar manner we can define the second Chern class of the spectral cover bundles as a vector, in the basis of the fiber class $F$ and classes pulled back from the base.
In other words, we can write 
\begin{eqnarray}
\nn
 c_2(TX) - c_2(V) &=& (W_B, a_F) \ ; \mbox{ where }\\
 W_B &:= &12 c_1(B) - \eta \ , \\
 \nn
a_F &:=& c_{2}(B)
   + c_{1}(B)^{2} \left(11+\frac{n^{3}-n}{24}\right)
   - \frac{n}{2}\left(\lambda^{2}-\frac{1}{4}\right)
\eta \cdot (\eta - n c_1(B)) \ .
\end{eqnarray}
Note that $a_F$ is some integer while $W_B$ is a curve class.
Thus a good norm to take is
\begin{equation}
|c_2(TX) - c_2(V)| = \sqrt{a_F^2 + W_B \cdot W_B} \in \IZ_{>0} \ ,
\end{equation}
where $W_B \cdot W_B$ is understood as curve intersection on the base and the fact that this is strictly greater than zero is the content of Proposition \ref{prop-spec} \footnote{In the corresponding plots, this manifests as the `blank' repeller around the x-axis.}.

\begin{figure}[!h!t!b]
\centerline{
(a)
\includegraphics[trim=0mm 0mm 0mm 0mm, clip, width=2.0in]{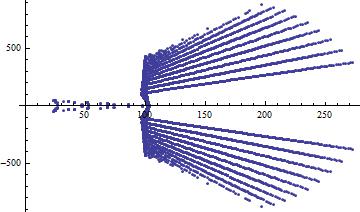}
(b)
\includegraphics[trim=0mm 0mm 0mm 0mm, clip, width=2.0in]{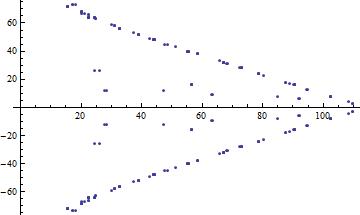}
(c)
\includegraphics[trim=0mm 0mm 0mm 0mm, clip, width=2.0in]{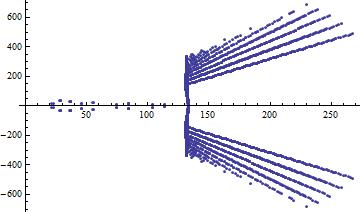}
}
\caption{{\sf {\small
A plot of $\ch_3(V)$ in the ordinate versus $| \ch_2(TX)_r - \ch_2(V)_r |$ in the abscissa for the 14264 $SU(n)$ spectral cover bundles on the elliptic fibrations over the 13 Hirzebruch surfaces $\IF_{r=0,\ldots,12}$ for $n=3,4,5$, shown respectively in (a), (b) and (c).
}}
\label{f:Fr}}
\end{figure}

Shown first are the plot of results for all the elliptic fibrations over the 13 Hirzebruch surfaces $\IF_{r=0,\ldots,12}$, there are a total of 14264 stable bundles thereon; these are shown in Figure \ref{f:Fr}.

\begin{figure}[!h!t!b]
\centerline{
(a)
\includegraphics[trim=0mm 0mm 0mm 0mm, clip, width=2.0in]{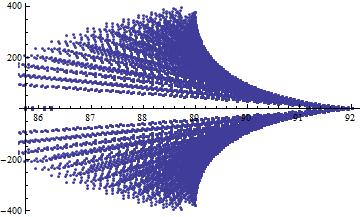}
(b)
\includegraphics[trim=0mm 0mm 0mm 0mm, clip, width=2.0in]{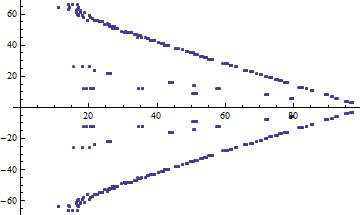}
(c)
\includegraphics[trim=0mm 0mm 0mm 0mm, clip, width=2.0in]{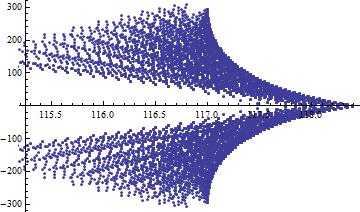}
}
\caption{{\sf {\small
A plot of $\ch_3(V)$ in the ordinate versus $| \ch_2(TX)_r - \ch_2(V)_r |$ in the abscissa for the 42352 $SU(n)$ spectral cover bundles on the elliptic fibrations over the 4 blow ups $\widehat{\IF}_{r=0,\ldots,3}$ of the Hirzebruch surfaces for $n=3,4,5$, shown respectively in (a), (b) and (c).
}}
\label{f:Frhat}}
\end{figure}

A similar situation is obtained for the elliptic threefolds over the 4 blowups of Hirzebruch surfaces, $\widehat{\IF}_{r=0,\ldots,3}$, there are a total of $42352$ stable $SU(n)$ bundles for $n=3,4,5$. These are shown in Figure \ref{f:Frhat}.

It is curious that in both classes of elliptically fibered threefolds, we find an alternating pattern of the distribution of Chern characters with respect to the parity of the bundle's rank, namely even and odd cases.
Although not shown in the plots, this pattern regarding parity persists at both lower (i.e., rank 2) and higher ranks of the bundle.

%%%%%%%%%%%%%%%%%%%%%%
\subsection{Cyclic Manifolds: Picard Rank $1$}
%Inspired by the statistics and plots of Chern classes in \cite{Marino:1998tb}, let us try to plot second versus third Chern classes of our manifolds and bundles in a meaningful way (the first Chern classes, of course, vanish for all our cases).

Now, the second Chern classes in all our cases, as discussed above, are vectors in the basis of the K\"ahler cone, this implies immediately, for $h^{1,1}(X)= 1$, or {\em cyclic}, Calabi-Yau threefolds, they are characterized by a single number.
From the readily available databases, we find there are 9 of such manifolds, 5 from the CICY list and 5 from the Kreuzer-Skarke hypersurface in toric fourfolds list, with the quintic at the intersection. For these threefolds we can visualize the solution space to DRY's sufficient conditions. 

The topological data for these 10 manifolds are conveniently summarized in Table 3 of \cite{Anderson:2007nc} and Eq (3.12) of \cite{He:2011rs}.
Though unfortunately we do not have enough data to draw any conclusions, for what we do have, we can readily plot the $c_2(TX)$ versus the Euler character $\chi(X) = c_3(TX)$; for clarity we record the values as well as the intersection form which here is just a number $J^3 = d$:
\begin{equation}
\begin{array}{ccc}
\begin{array}{|c|c|c|}\hline
c_2(TX) & \chi(X) & d \\
\hline\hline
 10 & -200 & 5 \\ \hline
 7 & -176 & 8 \\ \hline
 6 & -144 & 9 \\ \hline
 5 & 144 & 12 \\ \hline
 4 & -128 & 16 \\ \hline
 10 & -40 & 1 \\ \hline
 14 & -204 & 3 \\ \hline
 34 & -288 & 1 \\ \hline
 22 & -296 & 2 \\ \hline
\end{array}
& &
\begin{array}{l}
\includegraphics[trim=0mm 0mm 0mm 0mm, clip, width=3.5in]{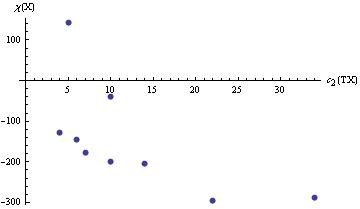}
\end{array}
\end{array}
\end{equation}

Now, for the DRY conjecture, the ample class can be parameterized by a single constant,  $H = a J$ for the single K\"ahler class $J$ and $a \in \IR_{>0}$.
Thus the DRY constraint simply gives
\begin{equation}
c_2(V) = n a^2 d + \frac{1}{24} c_2(TX) \ , \quad
c_3(V) < \frac{16 \sqrt{2}}{3} n a^3 d
\end{equation}
which by solving for $a$ leads the inequality
\begin{equation}
c_3(V) < \frac{(24 c_2(V) - n c_2(TX))^{3/2}}{9 \sqrt{3 n d}} \ .
\end{equation}
 {Notice this expression has no dependence on the underlying threefolds.}
Upon substituting the realistic values of the Chern classes for the cyclic threefolds, we find the following allowed regions in the $(c_2(V), c_3(V))$ plane, for $n=3,4,5$ respectively:
\[
\includegraphics[trim=0mm 0mm 0mm 0mm, clip, width=2in]{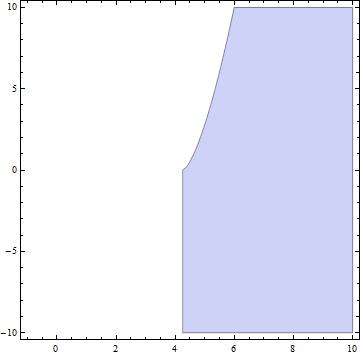}
\includegraphics[trim=0mm 0mm 0mm 0mm, clip, width=2in]{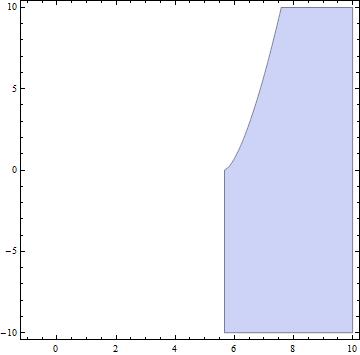}
\includegraphics[trim=0mm 0mm 0mm 0mm, clip, width=2in]{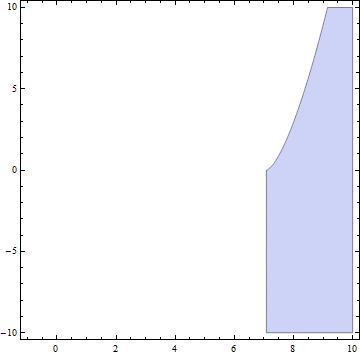}
\]

%%%%%%%%%%%%%%%%%%%%%%%%
\subsection{Picard Rank $2$ CICYs}
We can generalize the analysis of the previous section by going 1 step higher in $h^{1,1}$ and still permit ourselves to visualize.
Let us focus on favourable CICY manifolds with $h^{1,1} = 2$.
There are 36 of such spaces, including the famous bi-cubic in $\IP^2 \times \IP^2$.
Now, the second Chern class is conveniently encoded in a 2-vector, for example for the bi-cubic $\left[\begin{array}{c|cc}
\IP^{2} & 3 \\
\IP^{2} & 3 
\end{array}\right]$ with $(h^{1,1},h^{2,1}) = (2,83)$, 
we have two K\"ahler classes $(J_1,J_2)$, so that $c_2(TX) = (36,36)$ and the triple intersection form is $P(t_1,t_2) = 3 t_1^2t_2 + 3 t_1 t_2^2$.
We can therefore collect the triple $(c_2(TX)_1, c2(TX)_2, \chi(X))$ and plot in a three-dimensional space as follows:
\[
(a)
\includegraphics[trim=0mm 0mm 0mm 0mm, clip, width=4.0in]{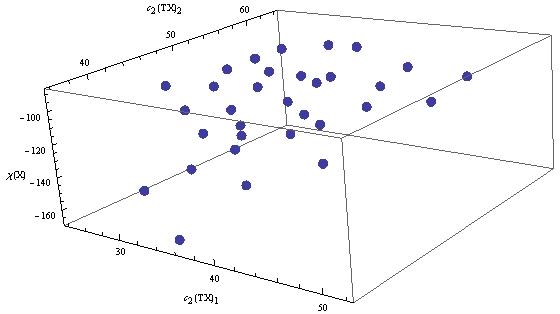}
(b)
\includegraphics[trim=0mm 0mm 0mm 0mm, clip, width=2.0in]{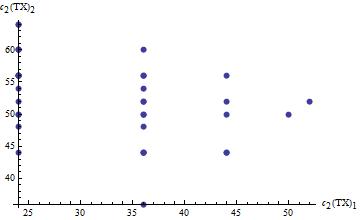}
\]
For clarity in part (b) of the above plot, we look at the vertical projection and see how the 2-vectors of the second Chern classes are distributed.

Now, the DRY conjecture will involve two parameters because we can take the ample class to be $H = a J_1 + b J_2$ with $a,b \in \IR_{>0}$.
Specifically, the conjecture becomes:
\begin{align}
\nn
\left[c_2(V) - \frac{n}{24} c_2(TX)\right]_{i=1,2} &= n
\left(\begin{array}{l}
a^2 d_{111} + 2ab d_{112} + b^2 d_{122} \\
a^2 d_{112} + 2ab d_{122} + b^2 d_{222}
\end{array}\right) \ , \\
c_3(V) &< \frac{16\sqrt{2}}{3}\ n (a^3 d_{111} + b^2 d_{222} + 3a^2bd_{112} + 3ab^2 d_{122}) \ ,
\end{align}
where the first two equations come from the 2 components of the second Chern class and $d_{ijk}$ as usual are the triple intersection numbers.

Again, we can use the first pair of equations to eliminate the parameters $(a,b)$ and substitute, for fixed $n=3,4,5$, into the inequality.
This time, there is no longer a common locus amongst {\it all} 36 manifolds which satisfy the inequalities; but there is no reason to expect this to be the case. What we can do is to plot for a particular manifold, the region of solutions to DRY, for $n=3,4,5$ respectively.
Let us choose, for concreteness, the bi-cubic:
\[
(a)
\includegraphics[trim=0mm 0mm 0mm 0mm, clip, width=2.0in]{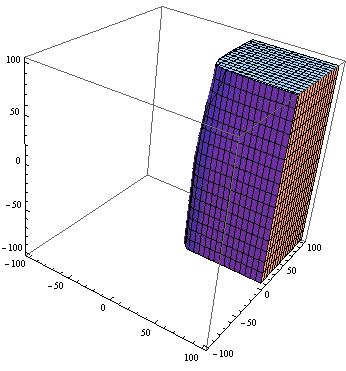}
(b)
\includegraphics[trim=0mm 0mm 0mm 0mm, clip, width=2.0in]{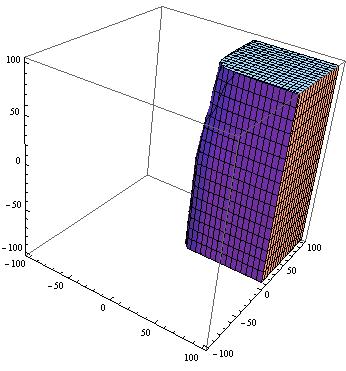}
(c)
\includegraphics[trim=0mm 0mm 0mm 0mm, clip, width=2.0in]{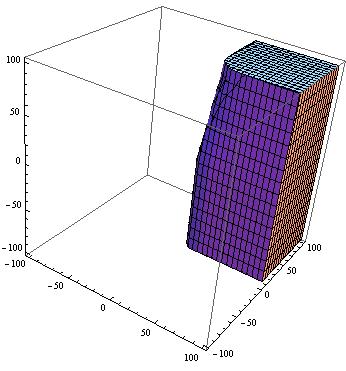}
\]

%%%%%%%%%%%%%%%%%%%%%%%%%%^^^^^^^^^^^^^^^^^^^^^^^^^^^^^^^^^^^^^
\subsection{General Spectral Cover Bundles}\label{scbstat}
Given the interesting pattern within the class of physically constrained spectral cover bundles,  it seems interesting to relax the physical bound on $c_2(V)$, and see whether the distribution of stable bundles have a different trend. We focus on one elliptically fibred Calabi-Yau threefold for concreteness,  taking the elliptic fibration over $\IF_1$.

Now, dropping the anomaly cancellation condition %as well as the condition that the the third Chern class be restricted to be multiples of 3, both of which come from physical motivations, 
we investigate the fibre component $a_F$ of $c_2(V)$ versus the possible $c_3(V)$.
The latter is no longer necessarily bounded in numerical value, we  consider the statistic of the following variables 
\begin{align}
\nn
x &= a_F(c_2(V)) = c_{2}(B)
   + c_{1}(B)^{2} \left(11+\frac{n^{3}-n}{24}\right)
   - \frac{n}{2}\left(\lambda^{2}-\frac{1}{4}\right)
\eta \cdot (\eta - n c_1(B))
\\ 
y &= c_3(V) =  2 \lambda \eta \cdot \left(\eta - n c_1(B) \right)
\end{align}
A different kind of statistics was considered earlier in \cite{Gabella:2008id} . 

Shown here are the plot of results for the $SU(5)$ bundle obtained in spectral cover construction, on all the elliptic fibrations over the 13 Hirzebruch surfaces $\IF_{r=0,\ldots,12}$. Comparing with Figure \ref{f:Fr}, we observe some additional features which became more obvious due to the relaxation of anomaly conditions on the Chern characters.

For example, it became clear there is a `banding' structure, which is in fact due to the increasing of the value of the parameter $\lambda$. 
\[
\includegraphics[trim=0mm 0mm 0mm 0mm, clip, width=3.6in]{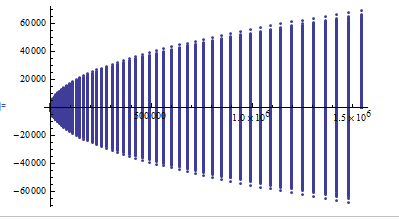}
\]
Moreover, it can be seen that the parabola-like shape is consistent with general predictions on numerical boundedness for stable bundles with specified rank and first Chern class. 

%%%%%%%%%%%%%%%%%%%%%%
\subsection{The Statistics of Generations}

Finally, it is interesting to look at the distribution of the number of generations, $\ch_3(V)$.
Of course, without constraints on the second Chern class, this number could vary freely.
We therefore impose the anomaly cancellation condition (allowing M5-branes), which gives an upper bound to the value of the second Chern class.
Thus, requiring that the second class be bounded above by $c_2(TX)$, we can plot the incidents of the possible $\ch_3(V)$ values.
Illustrated below are  the 13 elliptic fibrations over the Hirzebruch surfaces (in part (a) as well as the 4 over the blowups of these surfaces (in part (b)):
\begin{equation}
(a)
\includegraphics[trim=0mm 0mm 0mm 0mm, clip, width=3in]{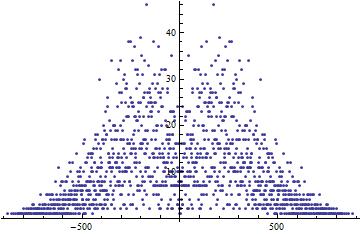}
(b)
\includegraphics[trim=0mm 0mm 0mm 0mm, clip, width=3in]{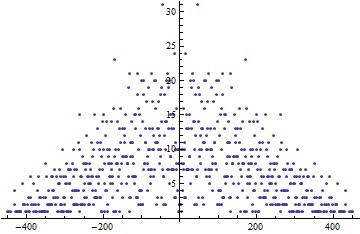}
\end{equation}
Of these, recall from the Proposition \ref{prop-spec} that there are none which saturate the bound by having $c_2(V) = c_2(TX)$.

%%%%%%%%%%%%%%%%
\section{Conclusions and Outlook}\label{concl}

In this paper, motivated by both physical and mathematical considerations, we explored a variety of constructions for stable vector bundles on Calabi-Yau threefolds, including spectral cover construction on elliptic fibrations, generalized Hartshorne-Serre constructions and monads. We have tried to understand especially the distribution of Chern numbers for stable and extremal vector bundles and the related question of anomaly cancellation in heterotic string theory. 
The construction of extremal bundles allows one to discuss the bundles fully within the perturbative heterotic string theory. A number of constructions mentioned here can yield extremal bundles ($c_1(V)=0$ and $c_2(V)=c_2(TX)$), but no systematic way to achieve this was known. 

In particular, in Section \ref{scbe} we showed, by enumerating all cases, that spectral cover bundles on elliptic fibration Calabi-Yau threefolds do not lead to extremal bundles. 
In the CICY poly-stable (direct sum) examples we find no phenomenologically interesting bundle which is actually extremal. 
In the positive monad case, we find more (cf.~Section (\ref{posmon}) for a summary). 
They are overall rare and indeed {\it very rare} when we impose also conditions on the value of $c_3(V)$. 

To remedy this situation, we have introduced a new method to obtain extremal bundles using a generalization of the Hartshorne-Serre construction. 
This construction (see (\ref{hs1}) and (\ref{hs2})) yields automatically locally free sheaves (i.e., vector bundles) and is advantageous in this sense over the usual Serre construction where the extension may not be even torsion free on a threefold. 
Another advantage of this method is that rank is practically unbounded and can indeed be naturally made very large. 

On elliptically fibered Calabi-Yau threefolds we study geometric features of these bundles but leave a full account of its stability to a separate publication.
Such a discussion will be based on a detailed understanding of moduli space of stable bundles and we feel such discussions is to some extent lacking in the current literature on heterotic string calculations. 
Another nice feature of the generalized Hartshorne-Serre construction is its close relation to the geometry of algebraic curves on threefolds. The class of curves associated to the Serre construction are usually called subcanoncial and understanding subcanonical curves on threefolds is generally a difficult problem. 
In the generalized setting, the curves need only be effective and this gives freedom in constructing bundles with physically desirable Chern numbers. We also discuss its relevance in proving existence statements in the DRY conjecture (see Section \ref{c3range}). 

%For the same reason, we may bring this class of bundles to bear on the DRY conjecture \cite{Douglas:2006jp}. Bundle and sheaf stability on algebraic surfaces has been well understood but the threefold case remains challenging. The DRY conjecture is motivated by the beautiful physical ideas of the split attractor flow and it predicts the existence of stable reflexive sheaves when certain relations among the Chern numbers are satisified by the bundle. 

We have made a few new observations regarding DRY (see Section \ref{DRYconj}). 
Realizing that on both favorable CICY and elliptic fibrations one can parameterize the ample divisors explicitly, we have found explicit solutions to DRY within the database of currently known stable vector bundles constructed from positive monads of rank $r\le 5$.
As observed at the end of section (\ref{exDRY}) there are some interesting exceptions, all of which are on self-Mirror Calabi-Yau threefolds, including the famous Schoen manifold. 
This populates the DRY range of the space of stable vector bundles. It should be noted that DRY predicts reflexive sheaves which contains vector bundles as a subset, while monads are necessarily locally free by construction.

In Section \ref{bbef} (especially \ref{bbDRY}) we studied both the Bogomolov and the DRY bounds on Chern numbers. The calculations quickly become impractical with larger Picard number and we restricted ourselves to the cases with Picard number does not exceed $2$. Here, by directly calculating the Chern numbers of the spectral cover bundles we see explicitly that they do violate DRY bounds in certain regions of the choice of the spectral cover curve. 
This is consistent with that the spectral cover bundles can only be shown to be stable with respect to special choices of polarizations. 
However, we should not conclude that they cannot in fact be stable for all polarizations.

The last part of our paper, Section \ref{chernstat}, contains various statistical results on Chern number distributions of stable vector bundles constructed using the aforementioned methods. 
Due to limitations of our data in terms of rank we can only see part of the grand picture. 
However, already we see quite interesting features, especially in the spectral cover construction where more data is known. 
Section \ref{scebstat} gives further results on extremal bundles and Section \ref{scbstat}, general spectral cover bundles. 
The results are consistent with a polynomial relation between $c_2(V)$ and $c_3(V)$, and appears sensitive to rank of $V$.
Although the plots contain vector bundles on different elliptically fibered Calabi-Yau threefolds the attractiveness of the patterns suggest they could be all connected together by singular deformations of the threefolds. Generally, it is our feeling that numerical methods may lead to 
further insights and effective bounds on the space of stable vector bundles which are very hard to obtain theoretically. 

%%%%%%%%%%%%%%%
\section*{Acknowledgements}

We are grateful to Lara Anderson, Ron Donagi, James Gray, An Huang, Babak Haghighat, Klaus Hulek, Tony Pantev, Michel van Garrel, Johannes Walcher, Baosen Wu, Noriko Yui for discussions. The first and third authors would like to thank Taida Institute for Mathematical Science, National Taiwan University for hospitality. PG thanks CRM Montr${\rm\acute e}$al, the Fields Institute, for hospitality where part of the work was carried out.  
YHH would like to thank the Science and Technology Facilities Council, UK, for grant ST/J00037X/1,
the Chinese Ministry of Education, for a Chang-Jiang Chair Professorship at NanKai University as well as the City of Tian-Jin for a Qian-Ren Scholarship, the US NSF for grant CCF-1048082, as well as City University, London and Merton College and Department of Theoretical Physics, Oxford, for their enduring support.
Moreover, he is indebted to the kind hospitality of his co-authors at the Department of Mathematics of Harvard University where this work began.

\newpage

%%%%%%%%%%%%%%%%%%%%%%%%%%%%=======================================

\end{document}